\documentclass[12pt]{article}
\pdfoutput=1
\usepackage[a4paper]{geometry}
\usepackage{jheppub, amsmath,amssymb,amsfonts,amsxtra, mathrsfs, makeidx,graphics,graphicx,amsthm,epsfig, ytableau,bm,longtable,float, color,tikz,mathtools,xfrac,footnote,rotating, lscape, makecell, environ,mathtools, empheq}

\usetikzlibrary{positioning}
\usetikzlibrary{chains}
\usetikzlibrary{arrows,fit,decorations.pathreplacing}
\tikzstyle{every picture}+=[remember picture]
\tikzstyle{na} = [baseline]

\usetikzlibrary{arrows, decorations.markings, calc, fadings, decorations.pathreplacing, patterns, decorations.pathmorphing, positioning}

\tikzstyle{every picture}+=[remember picture]
\tikzstyle{na} = [baseline=-.5ex]

\usepackage{bbm}
\usepackage{subfigure}

\newcommand{\eg}{\textit{e.g.}}

\newcommand{\ie}{\textit{i.e.}}

\numberwithin{equation}{section}

\newcommand{\nn}{\nonumber}

\newcommand{\be}{\begin{equation}} \newcommand{\ee}{\end{equation}}
\newcommand{\bea}{\begin{equation} \begin{aligned}} \newcommand{\eea}{\end{aligned} \end{equation}}

\def\tilde{\widetilde}

\def\hat{\widehat}

\def\a{\alpha}
\def\b{\beta}
\def\g{\gamma}

\def\rt2{\sqrt{2}}

\def\d{\partial}

\def\det{\mathop{\rm det}}

\def\CN{{\cal N}}

\def\CX{{\cal X}}

\def\1{{\ds 1}}

\def\repa{\raise4pt\hbox{$\square$}\mkern-14mu\raise-4pt\hbox{$\square$}}
\def\repab{\overline{\raise4pt\hbox{$\square$}\mkern-14mu\raise-4pt\hbox{$\square$}\mkern-1mu}}

\def\smileface{\ensuremath{\hbox{\large$\bigcirc$}\mkern-15mu\raise-1pt\hbox{\scriptsize$\smallsmile$}%
\mkern-10mu\raise4pt\hbox{..}\mkern4mu}}
\def\frownface{\ensuremath{\hbox{\large$\bigcirc$}\mkern-15mu\raise-1pt\hbox{\scriptsize$\smallfrown$}%
\mkern-10mu\raise4pt\hbox{..}\mkern4mu}}

\newcommand{\ba}{\begin{array}}
\newcommand{\ea}{\end{array}}
\newcommand{\bi}{\begin{itemize}}
\newcommand{\ei}{\end{itemize}}
\def\vec#1{\bm{#1}}
\def\bea#1\eea{\allowdisplaybreaks \begin{align}#1\end{align}}
 \newcommand{\ben}{\begin{enumerate}}
\newcommand{\een}{\end{enumerate}}
\newcommand{\bean}{\begin{eqnarray*}}
\newcommand{\eean}{\end{eqnarray*}}
\newcommand{\eref}[1]{(\ref{#1})}

\newcommand{\PE}{\mathop{\rm PE}}

\newcommand{\BZ}{\mathbb{Z}}

\newcommand{\comment}[1]{}

\definecolor{light-gray}{gray}{0.7}

\newcommand{\blue}{\color{blue}}

\def\aup#1 {\overset{#1}{\uparrow} \, \overset{\tilde{#1}}{\downarrow}}

\title{\centering{New 3d $\CN=2$ Dualities \\ from Quadratic Monopoles}}
\author[a]{Antonio Amariti,}
\author[b,c]{Ivan Garozzo,}
\author[b,d]{and Noppadol Mekareeya}

\affiliation[a]{INFN, Sezione di Milano, Via Celoria 16, I-20133 Milano, Italy}
\affiliation[b]{INFN, sezione di Milano-Bicocca, \\Piazza della Scienza 3, I-20126 Milano, Italy}
\affiliation[c]{Dipartimento di Fisica, Universit\`a di Milano-Bicocca, \\ Piazza della Scienza 3, I-20126 Milano, Italy}
\affiliation[d]{Department of Physics, Faculty of Science, \\
Chulalongkorn University, Phayathai Road, \\
Pathumwan, Bangkok 10330, Thailand}

\emailAdd{antonio.amariti@mi.infn.it}
\emailAdd{ivangarozzo@gmail.com}
\emailAdd{n.mekareeya@gmail.com}

\abstract{Aspects of three dimensional $\mathcal{N}=2$ gauge theories with
monopole superpotentials and their dualities are investigated. The moduli spaces of a number of such theories are studied using Hilbert series.
Moreover, we propose new dualities involving quadratic powers for the monopole superpotentials, for unitary, symplectic and orthogonal gauge groups.
These dualities are then tested using the three sphere partition function and matching of the Hilbert series.
We also provide an argument for the obstruction to the duality for theories with quartic monopole superpotentials.
}

\begin{document}

\maketitle

\section{Introduction}

The three dimensional $\mathcal{N}=2$ dualities is a classic subject \cite{deBoer:1997ka, deBoer:1997kr, Aharony:1997bx, Karch:1997ux, Aharony:1997gp} and remains to be an active field of research for several reasons.  This is partly because of its connection
with four dimensional dualities, as the Seiberg duality \cite{Seiberg:1994pq} and the S-duality.  
It is also because of the recent interest on 3d theory with lower supersymmetry, namely $\mathcal{N}=1$
(see \eg~ \cite{Armoni:2017jkl,Bashmakov:2018wts, Benini:2018umh,Gaiotto:2018yjh,Benini:2018bhk}),
and on the non supersymmetric case (see \eg~
\cite{Giombi:2011kc,Aharony:2011jz,Aharony:2012nh,Aharony:2015mjs,Seiberg:2016gmd}). 
One of the main reasons why the web of 3d $\mathcal{N}=2$ dualities is 
rich and interesting is related to the presence of a Coulomb branch, that corresponds to the moduli space of a dynamical
real scalar inside the vector multiplet.
The Coulomb branch can be parameterized in terms of complex coordinates,
obtained by combining the real scalar discussed above with the dual photon. This is possible because the gauge symmetry on the Coulomb branch is broken to the maximal abelian torus.
The complex combinations obtained from the real scalars and the dual photons correspond to the insertion of monopole 
 operators in the path integral.
 
 It is possible to modify the superpotential by the contribution of such monopole operators, parameterized in terms of the Coulomb 
branch coordinates \cite{Polyakov:1976fu, Affleck:1982as, Aharony:1997bx, Aharony:1997gp, deBoer:1997kr, Dorey:1998kq, Benini:2011mf}.
These superpotential have been shown to play a crucial role in 3d 
$\mathcal{N}=2$ dualities.
For example they have been used in the circle reduction of 4d Seiberg duality to define new effective dualities on $S^1$ \cite{Aharony:2013dha, Aharony:2013kma}. Moreover, monopole superpotentials have been used as a tool to study other field theories, such as dualites among $SU(2)$ gauge theories with eight chiral doublets \cite{Dimofte:2012pd}, dualities involving $pq$-webs \cite{Benvenuti:2016wet}, T-brane theories \cite{Collinucci:2017bwv}, compactification of Argyres--Douglas theories\footnote{
We mean the compactification of 4d 
$\mathcal{N}=1$ theories conjectured to enhance to 
Argyres--Douglas theories in the IR
\cite{Maruyoshi:2016aim,Agarwal:2016pjo,Agarwal:2017roi}.} to three dimensions and their mirror theories \cite{Benvenuti:2017lle, Benvenuti:2017kud, Benvenuti:2017bpg, Aghaei:2017xqe}, mirror theories of 3d $\CN=2$ supersymmetric QCD (SQCD) with zero superpotential \cite{Giacomelli:2017vgk}.  Higher order monopole potentials also have some applications in condensed matter and statistical physics (see \eg\, \cite{Block:2013gpa}).

Recently, a number of new 3d dualities for $U(N_c)$ gauge theories 
with monopole superpotentials have been obtained in \cite{Benini:2017dud}.
In most of these dualities the monopole superpotentials appear
with a linear power. 
They modify the constraint between the global charges, breaking 
the otherwise generated axial and/or topological symmetry.
Their presence modifies the rank of the dual gauge group, with respect
to the case without such linear monopole superpotential deformations.
Moreover an interesting proposal appeared in 
\cite{Benini:2017dud}, regarding a duality between $U(N_c)$ gauge groups and quadratic monopole superpotentials.

In this paper we elaborate on this proposal.  The three main points are as follows.  First, we obtain new classes of
unitary,  symplectic and orthogonal/special orthogonal gauge groups with quadratic monopole superpotentials.  Secondly, we study the moduli space of a large class of theories with monopole superpotentials using the Hilbert series \cite{Cremonesi:2013lqa, Cremonesi:2015dja, Hanany:2015via, Cremonesi:2016nbo, Cremonesi:2017jrk}.  Thirdly,  we check our proposals matching the Hilbert series and showing 
the consistency of the RG flow to the IR.  This last step is done 
using localization on $S^3$ \cite{Jafferis:2010un, Hama:2010av, Hama:2011ea, Kapustin:2009kz, Kapustin:2010xq}.
The dualities that we mostly discuss in this paper are summarized 
in table \ref{tab:summary}.
\\
{\footnotesize
\begin{longtable}{|c|c|c|}
\hline
Theory $A$ & Theory $B$ & Ref. \\
\hline
\begin{tabular}{@{}c@{}} $U(N_c)$ with $N_f$ flv., \\ $W=X^+ +X^-$ \end{tabular} &\begin{tabular}{@{}c@{}} $U(N_f-N_c-2)$ with $N_f$ flv., \\ $N_f^2$ singlets $M$, $W=M \tilde{q} q+ \hat{X}^+ +\hat{X}^-$ \end{tabular} & \cite{Benini:2017dud} \\
\hline
\begin{tabular}{@{}c@{}} $U(N_c)$ with $N_f$ flv., \\ $W=(X^+)^2 +(X^-)^2$ \end{tabular} &\begin{tabular}{@{}c@{}} $U(N_f-N_c)$ with $N_f$ flv., \\ $N_f^2$ singlets $M$, $W=M \tilde{q} q+ (\hat{X}^+)^2 +(\hat{X}^-)^2$ \end{tabular} & \cite{Benini:2017dud} \\
\hline \hline
\begin{tabular}{@{}c@{}} $U(N_c)$ with $N_f$ flv., \\ $W=X^-$ \end{tabular} &\begin{tabular}{@{}c@{}} $U(N_f-N_c-1)$ with $N_f$ flv., a singlet $S^+$ \\ $N_f^2$ singlets $M$, $W=M \tilde{q} q+ \hat{X}^+ + S^+ \hat{X}^-$ \end{tabular} & \cite{Benini:2017dud} \\
\hline
\begin{tabular}{@{}c@{}} $U(N_c)$ with $N_f$ flv., \\ $W=(X^-)^2$ \end{tabular} &\begin{tabular}{@{}c@{}} $U(N_f-N_c)$ with $N_f$ flv., a singlet $S^+$ \\ $N_f^2$ singlets $M$, $W=M \tilde{q} q+ (\hat{X}^+)^2 + S^+ \hat{X}^-$ \end{tabular} & \begin{tabular}{@{}c@{}}  Sec. \\ \ref{sec:Xmsq}  \end{tabular}  \\
\hline \hline
\begin{tabular}{@{}c@{}} $U(N_c)_{\frac{k}{2}}$ with $(N_f,N_f-k)$  \\ fund/antifund, $W=X^-$ \end{tabular} &\begin{tabular}{@{}c@{}} $U(N_f-N_c-1)_{-\frac{k}{2}}$ with $(N_f,N_f-k)$ fund/antifund \\ $N_f(N_f-k)$ singlets $M$, $W=M \tilde{q} q+ \hat{X}^+$ \end{tabular} & \cite{Benini:2017dud} \\
\hline
\begin{tabular}{@{}c@{}} $U(N_c)_{\frac{k}{2}}$ with $(N_f,N_f-k)$ \\  fund/antifund, $W=(X^-)^2$ \end{tabular} &\begin{tabular}{@{}c@{}} $U(N_f-N_c)_{-\frac{k}{2}}$ with $(N_f,N_f-k)$ fund/antifund \\ $N_f(N_f-k)$ singlets $M$, $W=M \tilde{q} q+ (\hat{X}^+)^2$ \end{tabular} & \begin{tabular}{@{}c@{}}  Sec. \\ \ref{sec:XmsqchflvCS}  \end{tabular}  \\
\hline \hline
\begin{tabular}{@{}c@{}} $USp(2N_c)$ with $2N_f$ fund, \\ $W=Y$ \end{tabular} &\begin{tabular}{@{}c@{}} $USp(2(N_f-N_c-2))$ with $2N_f$ flv., \\ $N_f(2N_f-1)$ singlets $M$, $W=M \tilde{q} q+ \hat{Y}$ \end{tabular} & \cite{Aharony:2013dha} \\
\hline \hline
\begin{tabular}{@{}c@{}} $USp(2N_c)$ with $2N_f$ fund, \\ $W=Y^2$ \end{tabular} &\begin{tabular}{@{}c@{}} $USp(2(N_f-N_c-1))$ with $N_f$ flv., \\ $N_f(2N_f-1)$ singlets $M$, $W=M \tilde{q} q+ \hat{Y}^2$ \end{tabular} & \begin{tabular}{@{}c@{}}  Sec. \\ \ref{sec:sympY2}  \end{tabular}  \\
\hline \hline
\begin{tabular}{@{}c@{}} $O(N_c)$ or $SO(N_c)$ with $N_f$ fund, \\ $W=Y$ \end{tabular} &\begin{tabular}{@{}c@{}} $O(N_f-N_c)$ or $SO(N_f-N_c)$ with $2N_f$ flv., \\ $N_f(2N_f+1)$ singlets $M$, $W=M \tilde{q} q+ \hat{Y}$ \end{tabular} & \cite{Aharony:2013kma} \\
\hline 
\begin{tabular}{@{}c@{}} $O(N_c)$ or $SO(N_c)$ with $N_f$ fund, \\ $W=Y^2$ \end{tabular} &\begin{tabular}{@{}c@{}} $O(N_f-N_c+2)$ or $SO(N_f-N_c+2)$ with $N_f$ flv., \\ $N_f(2N_f+1)$ singlets $M$, $W=M \tilde{q} q+ \hat{Y}^2$ \end{tabular} & \begin{tabular}{@{}c@{}}  Sec. \\ \ref{sec:orthY2}  \end{tabular}  \\
\hline \hline
\caption{Summary of the dualities}
\label{tab:summary}
\end{longtable}%
}

The paper is organized as follows.
In  section \ref{sec:rev-and-hs} we review the known dualities with monopole superpotentials.
After this brief review the Hilbert series for theories with monopole superpotentials are computed.  For theories that are dual to each other, we also match their Hilbert series.  This is by itself a new result and it will be useful in the discussion of the
Hilbert series for the cases with quadratic monopole
superpotentials.
 In  section \ref{sec:quadratic}
we introduce the dualities with quadratic monopole superpotentials.  Most of the dualities proposed in this section are new.
 In  section \ref{sec:S3part} we 
 show how to use the 3d partition function 
as a  consistency check for the new dualities.
 In  section \ref{sec:HSquad} 
 we compute the Hilbert series 
for the new dualities that we are proposing and match them across the duality.
In section \ref{sec-conc} we provide an argument for the obstruction to the duality of quartic monopole superpotentials, as well as discuss other interesting aspects that
we do not cover in the paper but deserve further investigations.

\section{Dualities with linear monopole superpotentials}
\label{sec:rev-and-hs}
In this section, we consider theories with linear monopole superpotentials.

\subsection{Review}

Dualities with linear monopole superpotential have been obtained 
in \cite{Benini:2017dud} by studying the dimensional reduction of the four dimensional
Seiberg duality between the $USp(2 N_c)$ gauge theory with $2 N_f$ 
fundamentals $Q$, and the $USp(2(N_f-N_c-2))$ gauge theory with $2 N_f$ fundamentals $q$, an anti-symmetric meson $M = Q Q$, and superpotential $W = M q q$ \cite{Intriligator:1995ne}.
The reduction of this theory on $S^1$ was studied in \cite{Aharony:2013dha}, and the duality 
was shown to be preserved if a Kaluza-Klein (KK) monopole superpotential 
was added to both the phases. 
The 3d limit considered in \cite{Benini:2017dud} consists of a large positive 
shift of the real masses of $N_f$ fundamentals and a large negative shift
for the remaining $N_f$ fundamentals.
Furthermore a similar shift was considered for the real scalar $\sigma$ in
the vector multiplet, in both the electric and the magnetic theory.
The two shifts have an opposite sign in order to keep the duality in the
IR.
This construction led to a 3d duality between the $U(N_c)$ gauge theory with $N_f$ fundamental flavors and superpotential 
\begin{equation}
W = X^+ + X^-
\end{equation}
and the $U(N_f-N_c-2)$ gauge theory with $N_f$ fundamental flavors, $N_f^2$ singlets $M$ and superpotential 
\begin{equation}
W = M q \tilde q + \hat X^+ + \hat X^-
\end{equation}
The superpotential terms for the 
monopoles $X^\pm$ and $\hat X^{\pm}$ break the axial and the
topological symmetry and constraint the $R$-charge 
of the fundamentals. 

A similar duality was constructed in \cite{Benini:2017dud}, involving only one monopole
superpotential. This duality can be constructed from four dimensions as
well and it consists of the $U(N_c)$ gauge theory with $N_f$ fundamental flavors and superpotential 
\begin{equation}
W =  X^-
\end{equation}
and the $U(N_f-N_c-2)$ gauge theory with $N_f$ fundamentals, $N_f^2$ singlets $M$ and superpotential 
\begin{equation}
W = M q \tilde q + \hat X^+ + \hat X^- S^+
\end{equation}
where $S^+$ corresponds to the monopole $X^+$ of the 
electric theory, acting as a singlet in the dual phase.

Another duality considered in \cite{Benini:2017dud} was obtained from this 
case by turning on a large real mass for $k$ anti-fundamentals.
This flow needs a large shift for the FI as well. The final
result consists of a relation between the $U(N_c)_{\frac{k}{2}}$ with $N_f$ fundamentals $Q$, $N_a=N_f-k$
antifundamentals $\tilde Q$ and superpotential 
\begin{equation}
W =  X^-
\end{equation}
and the $U(N_f-N_c-1)_{-\frac{k}{2}}$ with $N_f$ fundamental { and $N_a$ antifundamentals}, $N_f^2$ singlets $M$ and superpotential 
\begin{equation}
W = \sum_{i=1}^{N_f} \sum_{j=1}^{N_a} M^i_j q_i \tilde q^j + \hat X^+ ~.
\end{equation}

Dualities with linear monopole superpotential have been constructed for
real gauge groups as well.
They consists of the circle reduction of four dimensional Seiberg duality 
for $USp(2N_c)$ and $O(N_c)$ gauge groups.
These dualities have been constructed in \cite{Aharony:2013dha}  for the symplectic case
and in \cite{Aharony:2013kma} for the orthogonal case.
In the second case different constructions were needed, depending on the global properties of the gauge group.
The linear monopole superpotential in these cases is associated to the 
KK monopole, which is constructed algebraically from the affine root 
of the $B_{n}$, $C_{n}$ and $D_{n}$ series.

\subsection{Matching the Hilbert series}
In this section, we study the moduli spaces and compute the Hilbert series of a number of gauge theories with linear monopole superpotentials.  Let us briefly discuss some general features of such theories.  Suppose that $V$ is one of the basic monopole operator of the theory\footnote{In the case of the Chern-Simons theory, this basic monopole operator must be neutral under the gauge symmetry.}.   By putting $V$ in the superpotential, say $W=V$, the $R$-charge of $V$ is fixed to $2$.  This also results in fixing the $R$-charge of the chiral fields in the theory.  { Next, we consider the gauge theory whose gauge group is left unbroken by the monopole operators; this is known as the ``residual theory'' (see \cite{Cremonesi:2013lqa, Cremonesi:2015dja} for more details of this notion).}  If $V$ is the only basic monopole operator in the theory, then the gauge group is left unbroken by the monopole operator and the residual gauge theory is the same as the original theory.  However, if there are other basic monopole operators, there remain Coulomb branches parametrised by those monopole operators that need to be analysed.   

Let us now discuss this in several examples below.

\subsubsection{The Aharony duality for unitary gauge groups}
As a warm-up, let us consider the Aharony duality, proposed in \cite{Aharony:1997gp}:
\paragraph{Theory $A$:} $U(N_c)$ gauge theory with $N_f$ flavours and $W=0$.
\paragraph{Theory $B$:} $U(N_f-N_c)$ gauge theory with $N_f$ flavours $q$ and $\tilde{q}$, $N_f^2$ singlets $M$, singlets $S^\pm$ and superpotential $W= M \tilde{q} q+ S^- V^+ + S^+ V^-$, where $V^\pm$ are the basic monopole operators in theory $B$.
\\~\\
Note that $S^\pm$ in theory $B$ are mapped to the basic monopoles operators in theory $A$.  In what follows, we discuss the moduli space and the Hilbert series of theories $A$ and $B$.  In the following, we analyse the moduli space and compute the Hilbert series of these theories.

\subsubsection*{Theory $A$}
The Hilbert series of theory $A$ was studied in detail in \cite{Cremonesi:2015dja}.  Let us review the computation briefly here.  By the Callias index theorem, The monopole flux takes the form $(m_1, 0, \ldots, 0, m_{N_c})$ with $m_1 \geq 0  \geq m_{N_c}$.  The residual theories are as follows\footnote{We omit decoupled pure $U(1)$ gauge factors from the residual theory.}:
\ben
\item $m_1=m_{N_c}=0$.  The residual theory is the whole $U(N_c)$ with $N_f$ flavours. 
\item $m_1> 0= m_{N_c}$. The residual theory is $U(N_c-1)$ with $N_f$ flavours.  
\item $m_1= 0> m_{N_c}$. The residual theory is $U(N_c-1)$ with $N_f$ flavours. 
\item $m_1> 0> m_{N_c}$. The residual theory is $U(N_c-2)$ with $N_f$ flavours;
\een
Adding up the mesonic Hilbert series of these residual theories weighted by the factors associated to bare monopole operators, we obtain the Hilbert series of the total moduli space of theory $A$ as follows:
\be
\begin{split}
&H^{(A)}  (t,\vec{u},\vec{v},y; r )  \\
&= H^U_{N_c,N_f} (t,\vec{u},\vec{v},y; r ) \\
& \quad+ \left( \sum_{m_1=1}^\infty t^{|m_1| P} z^{m_1} y^{m_1(- N_f) } + \sum_{m_{N_c}=-\infty}^{-1} t^{|m_{N_c}| P} z^{m_{N_c}} y^{-m_{N_c}(-N_f)} \right)  H^U_{N_c-1,N_f} (t,\vec{u},\vec{v},y; r )  \\ 
& \quad + \sum_{m_1=1}^\infty \sum_{m_{N_c} = -\infty}^{-1} t^{|m_1-m_{N_c}|P} z^{m_1+m_{N_c}} H^U_{N_c-2,N_f} (t,\vec{u},\vec{v},y; r )
\end{split}
\ee
and so
\be \label{HSAAharony}
\begin{split}
& H^{(A)}  (t,\vec{u},\vec{v},y; r ) \\
&= H^U_{N_c,N_f} (t,\vec{u},\vec{v},y; r )  + \left[ \frac{t^P y^{-N_f} z}{1- t^P y^{-N_f} z} +   \frac{t^P y^{-N_f} z^{-1}}{1- t^P y^{-N_f} z^{-1}}\right]H^U_{N_c-1,N_f} (t,\vec{u},\vec{v},y; r )   \\
& \qquad +  \left[\frac{t^P y^{-N_f} z}{1- t^P y^{-N_f} z}  \cdot \frac{t^P y^{-N_f} z^{-1}}{1- t^P y^{-N_f} z^{-1}}  \right] H^U_{N_c-2,N_f} (t,\vec{u},\vec{v},y; r ) 
\end{split}
\ee
where $r$ is the $R$-charge of the quarks, and the mesonic Hilbert series for $U(N_c)$ with $N_f$ flavours is given by \cite{Cremonesi:2015dja, Gray:2008yu}
\be \label{mesHU}
\begin{split}
& H^U_{N_c,N_f} (t,\vec{u},\vec{v},y; r ) \\
&= \sum_{n_1, \ldots, n_{N_c} \geq 0}[0^{N_f-N_c-1},n_{N_c},\ldots,n_1; n_1, \ldots, n_{N_c}, 0^{N_f-N_c-1}]_{\vec{u},\vec{v}} (t^r y)^{2\sum_{j=1}^{N_c} {j n_j}} 
\end{split}
\ee
with $y$ the fugacity for the $U(1)$ axial symmetry, and $P$ is the $R$-charge of the basic monopole operators:
\be
P = N_f(1-r) - (N_c-1)~.
\ee
The moduli space is generated by $X^+$, $X^-$ and the $N_f \times N_f$ meson matrix $M$, subject to the following relations
\be
\begin{split}
\epsilon^{i_1 i_2 \cdots i_{N_f}} \epsilon_{j_1 j_2 \cdots j_{N_f}}  M_{i_1} ^{j_1} \cdots M_{i_{N_c+1}} ^{j_{N_c+1}} &=0~, \\
X^\pm \epsilon^{i_1 i_2 \cdots i_{N_f}} \epsilon_{j_1 j_2 \cdots j_{N_f}}  M_{i_1} ^{j_1} \cdots M_{i_{N_c}} ^{j_{N_c}} &=0~,\\
X^+ X^- \epsilon^{i_1 i_2 \cdots i_{N_f}} \epsilon_{j_1 j_2 \cdots j_{N_f}}  M_{i_1} ^{j_1} \cdots M_{i_{N_c-1}} ^{j_{N_c-1}} &=0  ~.
\end{split}
\ee
The first set of relations implies that the rank of $M$ is at most $N_c$, the second set of relations implies that if $X^+ \neq 0$ or $X^- \neq 0$ the rank of $M$ is at most $N_c-1$, and the third set of relations implies that if both $X^+ \neq 0$ and $X^- \neq 0$ the rank of $M$ is at most $N_c-2$.

\subsubsection*{Theory $B$}
The rank of the gauge group of each residual theory in theory A put a restriction on the rank of the mesons.  Since the meson of theory $A$ is mapped to $M$ in theory $B$, we look at various possible rank of $M$ here.  Subsequently, we follow the argument of \cite{Aharony:1997gp}. 

If we give $M$ a vacuum expectation value (VEV) of rank $N_c$, the low energy theory consists of a $U(N_f-N_c)$ gauge group with $N_f-N_c$ flavours.  The latter has a dual description as the WZ theory with superpotential\footnote{Suppose that the $R$-charges of $q$ and $\tilde{q}$ are $R[q]=R[\tilde{q}]=R$.  In an effective theory of $U(N_f-N_c)$ gauge group with $N_f-N_c-p$ flavours, we have $R[V^\pm]=(N_f-N_c-p)(1-R)-(N_f-N_c-1)$ and $R[\det(q\tilde{q})] = 2R (N_f-N_c-p)$.  Hence, the combination $V^+ V^- \det(q \tilde{q})$ has $R$-charge $2(1-p)$.} $-V^+ V^- \det(q \tilde{q})$ \cite[(8.8)]{Aharony:1997bx}.  Hence the effective superpotential of theory $B$ in this case is 
\be W'_B = -V^+ V^- \det(q \tilde{q}) + S^+ V^- +S^- V^+~.\ee 
The $F$-terms $\partial_{V^\pm} W'_B =0$ imply that $ S^\mp  = V^\mp \det(q \tilde{q})$.  On the other hand, the $F$-terms $\partial_{S^\pm} W'_B =0$ imply that $V^\mp=0$.  Thus, $S^\pm=0$ in the chiral ring.  Since $S^\pm$ are mapped to the basic monopole operators of theory $A$, this case corresponds to case 1 of theory $A$, in which the monopole fluxes $m_1=m_{N_c}=0$.  Indeed, the Hilbert series of the space generated by an $N_f\times N_f$ matrix $M$ such that $\mathrm{rank}(M) \leq N_c$ is equal to $H^{U}_{N_c, N_f}$.

On the other hand, if we give $M$ a VEV of rank $N_c-1$, the low energy theory consists of a $U(N_f-N_c)$ gauge group with $N_f-N_c+1$ flavours.  The latter has a  low energy effective description as the WZ theory with superpotential $-2(V^+ V^- \det(q \tilde{q}))^{1/2}$ (see (8.6) and the discussion just above (8.7) of \cite{Aharony:1997bx}), which is valid away from the origin where $V^+=V^-=0$.  The effective superpotential of theory $B$ is therefore
\be W''_B = -2(V^+ V^- \det(q \tilde{q}))^{1/2} + S^+ V^- +S^- V^+~.\ee   The $F$-terms $\partial_{S^\pm} W''_B =0$ imply that $V^{\pm}=0$, \ie~ the Coulomb branch of theory $B$ is lifted.  The case in which $S^\pm =0$ corresponds to case 1 of theory $A$.    Away from the origin, the $F$-terms $\partial_{V^\pm} W''_B =0$ imply that
\be
-(V^+ V^- \det(q \tilde{q}))^{-1/2} V^\mp \det(q \tilde{q}) + S^\mp =0~,
\ee
or equivalently
\be \label{SpmAharony}
S^\pm = (V^{\pm} \det(q \tilde{q}))^{1/2} (V^\mp)^{-1/2}~.
\ee
We see that $S^\pm$ can take arbitrary values.  The case in which $S^+\neq0$ or $S^- \neq 0$ corresponds to cases 2 and 3 of theory $A$ in which the monopole fluxes $m_1>0, \, m_{N_c}=0$ or $m_1=0, \, m_{N_c}<0$.  { The Hilbert series of these cases corresponds to the second and the third term in the first line of \eref{HSAAharony}.}

If $M$ is given a VEV of rank $N_c-2$, the low energy theory consists of a $U(N_f-N_c)$ gauge group with $N_f-N_c+2$ flavours.  The effective superpotential of theory $B$ is $W'''_B = -3(V^+ V^- \det(q \tilde{q}))^{1/3} + S^+ V^- +S^- V^+$.  The $F$-terms $\partial_{S^\pm} W''_B =0$ imply that $V^{\pm}=0$.  The case in which $S^\pm =0$ corresponds to case 1 of theory $A$.  Away from the origin, using the $F$-terms $\partial_{V^\pm} W''_B =0$, we find that $S^\pm = \left( \frac{\det(q \tilde{q}) V^\pm}{(V^\mp)^2} \right)^{1/3}$.  Substituting these back to $W'''_B$, we obtain the effective superpotential $W'''_B = -( V^+ V^-  \det(q \tilde{q}))^{1/3}$.  Hence, $S^\pm$ can take arbitrary values.  Again, the case in which $S^+\neq0$ or $S^- \neq 0$ corresponds to cases 2 and 3 of theory $A$.  The case in which both $S^+$ and $S^-$ are non-zero corresponds to case 4 of theory $A$, in which $m_1 \neq 0$ and $m_{N_c} \neq 0$.  { The Hilbert series of the latter case corresponds to the second line of \eref{HSAAharony}.}

If the VEV of $M$ has rank greater than $N_c$, say $\mathrm{rank}(M)=N_c+p$ with $p>0$, the low energy theory consists of a $U(N_f-N_c)$ gauge group with $N_f-N_c-p$ flavours\footnote{Let $R[q] = R[\tilde{q}] = r$. Then $R[V^\pm] = (N_f-N_c-p)(1-r) - (N_f-N_c-1)$ and $R[ \det(q \tilde{q})] = 2r(N_f-N_c-p)$, and so the combination $V^+ V^- \det(q \tilde{q})$ has $R$-charge $2(1-p)$.}.   We will see that there is no stable supersymmetric vacuum for any $p>0$.
\bi
\item If $p>1$, from \cite[(8.6)]{Aharony:1997bx}, the effective superpotential is $(p-1)(V^+ V^- \det(q \tilde{q}))^{-\frac{1}{p-1}}$, and for the superpotential 
\be
W_B'= (p-1)(V^+ V^- \det(q \tilde{q}))^{-\frac{1}{p-1}} +  S^+ V^- +S^- V^+~.
\ee
Using the equations of motion, we obtain 
\be
W_B' \sim \left(\frac{S^+ S^-}{\det(q \tilde{q})} \right)^{\frac{1}{p+1}}
\ee 
and so we have runaway vacua in this case. 
\item If $p=1$, according to \cite{Aharony:1997bx}, this theory is described by the relation $V^+ V^- \det(q \tilde{q}) =1$. The effective superpotential of theory $B$ is therefore
\be
W_B' =  \lambda ( V^+ V^- \det(q \tilde{q})-1 ) +  S^+ V^- +S^- V^+~,
\ee
where $\lambda$ is a Lagrange multiplier.  Using the equations of motion, we can rewrite $W_B'$ as $W_B' \sim \left(\frac{S^+ S^-}{\det(q \tilde{q})}\right)^{\frac{1}{2}}$, and so we also have runaway vacua.
\ei

\subsubsection{The duality involving $W=X^+ + X^-$}
Let us consider the following pair of theories \cite{Benini:2017dud}:
\paragraph{Theory $A$:} $U(N_c)$ gauge theory with $N_f$ flavours and 
\be
W_A=X^+ + X^-~.
\ee
\paragraph{Theory $B$:} $U(N_f-N_c-2)$ gauge theory with $N_f$ flavours $q$ and $\tilde{q}$, $N_f^2$ singlets $M$ and superpotential 
\be W_B= M \tilde{q} q+ \hat{X}^+ +\hat{X}^-~.
\ee

In theory $A$, the flux of the monopole operator takes the following form $\vec{m}=(m_1, 0,\ldots,0, m_{N_c})$ with $m_1 \geq 0 \geq m_{N_c}$. Here $X^+$ denotes the monopole with the flux $(1,0,\ldots,0)$ and $X^-$ denotes the monopole with the flux $(0,\ldots,0,-1)$.   { The presence of $X^\pm$ as the linear terms in the superpotential $W_A$ sets $X^\pm=0$.} Hence, all monopole operators vanish in the chiral ring.  The Hilbert series of theory A is therefore the mesonic Hilbert series of $U(N_c)$ with $N_f$ flavours and zero superpotential:
\be
\begin{split}
H^{(A)} &= H^U_{N_c,N_f} (t,\vec{u},\vec{v}; r ) ~,
\end{split}
\ee 
where the $R$-charge $r$ of the quarks fixed by the superpotential $W_A$:
\be
N_f(1-r)-(N_c-1)=R[X_\pm]=2  \quad \Rightarrow \quad r=1-(N_c+1)/N_f~.
\ee
and the expression for $H^U_{N_c,N_f} (t,\vec{u},\vec{v}; r )$ is given by \eref{mesHU}.  Here there is no fugacity $y$ because the $U(1)$ axial symmetry is completely broken.

We compute the Hilbert series of theory $B$ as follows.  We first perform the Aharony dual to theory $B$ and obtain
\paragraph{Theory $B'$:} $U(N_c+2)$ gauge theory with $N_f$ flavours $\mathfrak{q}$ and $\tilde{\mathfrak{q}}$, two singlets $\hat{X}^\pm$ and superpotential
\be
W_{B'}=  \hat{X}^+V^- +\hat{X}^-V^++ \hat{X}^++\hat{X}^-~,
\ee
where $V^\pm$ are the basic monopoles in theory $B'$.  The $F$-terms $\partial_{\hat{X}^\pm} W_{B'}=0$ gives 
\be \label{Vmpm1} V^{\mp}=-1~.\ee  
This means that the monopole operators have non-zero fluxes $(\hat{m}_1, 0, \ldots, \hat{m}_{N_c+2})$ with $\hat{m}_1 \neq \hat{m}_{N_c+2} \neq 0$.  Thus, the gauge group $U(N_c+2)$ is broken to $U(N_c) \times U(1)^2$. Moreover, writing $V^\pm= e^{z_\pm}$ and considering the $F$-terms $\partial_{z_\pm} W_{B'}=0$, we find that $\hat{X}^\pm V^{\mp}=0$.  From \eref{Vmpm1}, we find that 
\be \hat{X}^\pm =0~,
\ee
\ie~ $\hat{X}^\pm$ vanish in the chiral ring. Thus, the residual theory is $U(N_c)$ gauge theory with $N_f$ flavours and zero superpotential.  The $R$-charge $R$ of the flavour fields $\mathfrak{q}$ and $\tilde{\mathfrak{q}}$ in theory $B'$ is given by
\be
N_f(1-R)-(N_c+2-1)= R[V^\pm]=0 \quad \Rightarrow \quad R= 1-(N_c+1)/N_f=r~.
\ee
Indeed, the Hilbert series of theory $B$, which is dual to theory $B'$, is given by
\be
H^{(B)} = H^U_{N_c,N_f} (t,\vec{u},\vec{v}; r=1-(N_c+1)/N_f ) = H^{(A)}~.
\ee

\subsubsection{The duality involving $W=X^-$} \label{sec:HSXm}
Let us now consider the following pair of theories \cite{Benini:2017dud}:
\paragraph{Theory $A$:} $U(N_c)$ gauge theory with $N_f$ flavours $Q$ and $\tilde{Q}$ and $W=X^-$.
\paragraph{Theory $B$:} $U(N_f-N_c-1)$ gauge theory with $N_f$ flavours $q$ and $\tilde q$, $N_f^2$ singlets 
$M$,
singlet $S^+$ and superpotential $W=M \tilde{q} q + \hat X^+ + \hat X^- S^+$ , where $\hat X^{\pm}$ are the basic monopoles in theory $B$, and $S^+$ is dual to the monopole $X^+$ in theory $A$.
\\

The first observation is that the monopoles in both theories do not appear in the superpotential in a symmetric way,
thus the charge conjugation is broken. Moreover, the $U(1)$ topological symmetry and the $U(1)$ axial symmetry are broken to a diagonal subgroup which we shall henceforth refer to as $U(1)_{T'}$.
As a consequence we need to slightly modify the expression for the 
monopole $R$-charge. 

\subsubsection*{Theory $A$}

Let us consider theory $A$ and define
\be
R = \frac{1}{2}(R[Q]+R[\tilde Q])~.
\ee
The $R$-charges of the $X^\pm$ are given by
\be \label{paramono}
\begin{split}
	R[X^-]-\a &=N_f(1-R)-(N_c-1) \\
	R[X^+]+\a &=N_f(1-R)-(N_c-1),
\end{split}
\ee
where $\a$ parametrizes the mixing of the $R$-charge and the $U(1)_{T'}$ symmetry.  { More explicitly, we consider the linear combination of the $U(1)_R = U(1)_{R_0}- \alpha U(1)_{T'}$, where $U(1)_{R_0}$ is the $R$-charge of $X^\pm$ before mixing and $T'[X^\pm] = \pm 1$.}   { Note also that this parametrisation is very similar to the one used in \cite{Giacomelli:2017vgk}.\footnote{On page 9 of \cite{Giacomelli:2017vgk}, the authors studied the $U(1)$ gauge theory with 2 flavours and the monopole superpotential $V^+ + {\cal X}_1 V^-$, where $\CX_1$ is a neutral chiral multiplet under the $U(1)$ gauge group. To avoid a potential confusion with the notation in this paper, let us denote by $a$ here the notation $\alpha$ on pages 8 and 9 of \cite{Giacomelli:2017vgk}.  From \cite[(2.8)]{Giacomelli:2017vgk}, the $R$-charges of quarks and antiquarks are given by $\frac{1-a}{2}$.  The $R$-charges of the monopole operators $V^\pm$ can be parametrised precisely as in \eref{paramono} as $R[V^+] =2 = 2 \left(  1- \frac{1-a}{2} \right) - \alpha$ and $R[V^-] = 2 \left(  1- \frac{1-a}{2} \right) + \alpha$.  Solving these two equations, we see that $\alpha = -1 +a$ and $R[V^-]=2a$; hence $R[{\cal X}_1] = 2- R[V^-]= 2- 2a$ in accordance with the discussion on page 9 of \cite{Giacomelli:2017vgk}.}} We shall soon show that this parametrisation is consistent with the proposed duality.  

The superpotential fixes the $R$-charge of $X^-$ to be $R[X^-]=2$, and so
\be \label{alphaNfNc}
	N_f(1-R)-(N_c-1)+\a=2,
\ee
from which we get $R$
\be
	R=\frac{N_f -N_c - 1 + \a}{N_f},
\ee
so the meson has $R$-charge
\be
	2 R=\frac{2(N_f -N_c - 1+\a)}{N_f}~.
\ee

Now let us turn our attention to the Hilbert series. However, we still have the Coulomb branch generated by the basic monopole operator $X^+$. Hence, to compute the Hilbert series we can use a similar argument to that of \cite{Aharony:2013kma}. We have two cases to analyse:
\bi
\item {\bf $m_1=0$:} with residual theory $U(N_c)$ with $N_f$ flavours and $W=0$. The Hilbert series 
is 
\be \label{casem1zero}
	H_{I}^{(A)}(t, \vec u, \vec v,z ; R)=H^U_{N_c, N_f}(t, \vec u, \vec v,z ; R)~,
\ee
where $z$ is the fugacity for the $U(1)_{T'}$ symmetry.
\item {\bf $m_1>0$:} with residual theory $U(N_c-1)$ with $N_f$ flavours and $W=0$. The Hilbert series is given by
\be \label{casem1nonzero}
	H_{II}^{(A)}(t, \vec u, \vec v, z; R)=\sum_{m_1=1}^{+\infty} t^{m_1 R[X^+]} z^{m_1 T'[X^+]} H^U_{N_c-1, N_f}(t, \vec u, \vec v, z; R),
\ee
where $z^x$ is the fugacity for the monopole operator $X^+$ with flux $(1,0,\ldots,0)$.  The sum gives
\be
	\sum_{m_1=1}^{+\infty} \left(t^{[N_f(1-R)-(N_c-1)]} z^x\right)^{m_1}=
	\frac{t^{[N_f(1-R)-(N_c-1)]} z^x}{1-t^{[N_f(1-R)-(N_c-1)]} z^x}.
\ee
\ei
Thus, the Hilbert series of theory $A$ reads
\be
\begin{split}
	H^{(A)}(t, \vec u, \vec v, z; R) &= ( H_{I}^{(A)}+ H_{II}^{(A)}) (t, \vec u, \vec v,z; R) \\
	&= H^U_{N_c, N_f}(t, \vec u, \vec v,z; R) + \frac{t^{[N_f(1-R)-(N_c-1)]} z^x}{1-t^{[N_f(1-R)-(N_c-2)]} z^x}H^U_{N_c-1, N_f}(t, \vec u, \vec v, z; R)~.
\end{split}
\ee

\subsubsection*{Theory $B$}
Let us analyze now the theory $B$. As we did for theory $A$, let us define 
\be
r = \frac{1}{2}(R[q]+R[\tilde q])~.
\ee 
For the monopoles, we parametrises by $\beta$ the mixing of the $R$-charge and the $U(1)_{T'}$ symmetry and obtain
\bea
	&R[\hat X^-]+\b=N_f(1-r)-(N_f - N_c -2) \\
	&R[\hat X^+]-\b=N_f(1-r)-(N_f - N_c -2). 
\eea
The superpotential imposes $R[\hat X^+]=2$, from which we get $r$
\be \label{rbeta}
	r=\frac{N_c + \b}{N_f}.
\ee
Using this expression for $r$ we find 
\be
	R[\hat X^-]=2(1-\b),
\ee
and also that 
\be \label{Spbeta}
	R[S^+]=2-R[\hat X^-]=2\b.
\ee
For the singlets $M$ we find
\be
	R[M]=2-2r=\frac{2(N_f-N_c-\b)}{N_f}.
\ee

Let us now relate the $R$-charges of various fields in theory $A$ to those in theory $B$.  In theory $B$, both $\hat{X}^+$ and $\hat{X}^-$ vanish in the chiral ring.  { The former is due to the presence of $\hat{X}^+$ in the superpotential}, whereas the latter follows from $\partial_{S^+} W=0$.  Moreover, $X^-$ in theory $A$ and $\hat{X}^+$ in theory $B$ both have $R$-charge $2$ due to the superpotentials.  We thus propose the following duality maps:
\be \label{opmapsinglinmono}
\begin{split}
\text{Theory $A$} &\qquad \text{Theory $B$} \\
X^+ \quad &\longleftrightarrow \quad S^+~, \\
\text{mesons} \quad &\longleftrightarrow \quad M~.
\end{split}
\ee
Due to the term $M \tilde{q} q$ in the superpotential, we also have
\be
R=1-r~.
\ee
The map between the mesons and $M$ implies that
\be \label{alphabetaeq}
\frac{2(N_f -N_c - 1+\a)}{N_f}=\frac{2(N_f-N_c-\b)}{N_f}   \quad  \implies \quad \a=1-\b~.
\ee
Equating $R[X^+]=R[S^+]$ and using \eref{rbeta}, \eref{Spbeta}, we find that\footnote{Let us compare the charge assignment here with that in \cite{Benini:2017dud}.  In the latter, $R[Q]$ and $R[\tilde{Q}]$ are chosen to be $R= \frac{N_f-N_c}{2N_f}$ and so $R[S^+] = N_f-N_c$.}
\be \label{RSplus}
	\b=r N_f -N_c =N_f(1-R)-N_c~, \qquad  R[S^+] = 2[N_f(1-R) -N_c]~.
\ee
Using \eref{alphabetaeq} and \eref{RSplus}, we find that $N_f(1-R)-(N_c-1)+\alpha=2$, which is consistent with \eref{alphaNfNc}.  This shows that our choice of parametrisation of the mixing between $U(1)_R$ and $U(1)_{T'}$ is consistent with the proposed duality.

In order to study the moduli space of theory $B$, we consider its Aharony dual, which is given by
\paragraph{Theory $B'$}: $U(N_c + 1)$ with $N_f$ flavours and singlets  $\hat X^\pm$ and $S^+$ with superpotential
$W_{B'}=\hat X^+ V^- + \hat X^- V^+ + \hat X^+ + \hat X^- S^+$, where $V^\pm$ are the basic monopole operators in this theory.
\\~\\
Consider first the $F-$terms obtained by 
differentiating with respect to the singlets:
\bea
	&\d_{\hat X^-} W_{B'}=0 \quad \implies \quad V^+=-S^+, \label{VpSplin} \\
	&\d_{S^+} W_{B'}=0 \quad \implies \quad \hat X^-=0, \\
	&\d_{\hat X^+} W_{B'}=0 \quad \implies \quad V^-=-1.
\eea 
Plugging these equations into $W_{B'}$ we get 
\be
	W_{B'}=0.
\ee
Observe the following (it will be needed later): $V^-=-1$ implies that the flux $m_{N_c+1} \ne 0$, hence the gauge group
breaks to $U(N_c+1) \to U(N_c) \times U(1)$; moreover $V^+=-S^+$ implies that we need to consider two cases for the 
computation of the Hilbert series, namely $S^+=0$ and $S^+ \ne 0$.  We shall see below that these correspond to the cases of $m_1=0$ and 
$m_1 \ne 0$ in the theory $A$, namely \eref{casem1zero} and \eref{casem1nonzero} respectively. 

Let us turn our attention to the $R$-charges.  Let $\mathfrak q$ and $\tilde{\mathfrak q}$ be the fundamentals and antifundamentals of theory $B'$.  Let us also define $R_{B'}=\tfrac{1}{2}(R[\mathfrak q] + R[\mathfrak {\tilde q}])$.
The superpotential $W_{B'}$ fixes the $R$-charge of $\hat X^+$ to be $R[\hat X^+]=2$, and so $R[V^-]=0$.

The $R$-charges of the monopole operators $V^\pm$ of theory $B'$ is given by
\bea
	R[V^+]&=N_f(1-R_{B'})-N_c + \g \\
	R[V^-]&=N_f(1-R_{B'})-N_c - \g =0 .
\eea
where $\g$ parametrise the mixing between the $U(1)$ $R$-symmetry and the $U(1)_{T'}$ symmetry.  We thus have
\be	
	\begin{cases}
		N_f(1-R_{B'})-N_c - \g= R[V^-] =0~ ,  \\
		N_f(1-R_{B'})-N_c +\g = R[V^+] \overset{\eref{VpSplin}}{=} R[S^+] \overset{\eref{RSplus}}{=}  2[N_f(1-R) -N_c].
	\end{cases}
\ee
The solution for $R_{B'}$ reads
\be
	R_{B'}=R~,
\ee
and so 
\be
\gamma = 1-\alpha~.
\ee
Hence, the $R$-charge of the mesons in this theory is $R[M]=2R_{B'}=2R$, which perfectly match with the $R-$charge of the singlets $M$ in theory $B$. 

Now we are ready to compute the Hilbert series. As we said, we have two cases: for $S^+=0$ we have $U(N_c)$ with $N_f$ flavours and $W_{B'}=0$, so
\be
	H_I^{(B')}(t, \vec u, \vec v, z; R_{B'})=H^U_{N_c, N_f}(t, \vec u, \vec v, z; R_{B'}).
\ee
When $S^+ \ne 0$ the gauge group breaks to $U(N_c - 1)$ with a dressing factor due to the presence of $V^+$; thus
the Hilbert series is 
\be
	H_{II}^{(B')}(t, \vec u, \vec v, z; R_{B'})=\sum_{m_1=1}^{+\infty} t^{m_1 R[V^+]} z^{m_1 T'[V^+]} H^U_{N_c-1, N_f}(t, \vec u, \vec v, z; R_{B'}),
\ee
again, as theory $A$ we find
\be
	H_{II}^{(B')}(t, \vec u, \vec v, z; R_{B'})=\frac{t^{[N_f(1-R_{B'})-(N_c-1)]} z^x}{1-t^{[N_f(1-R_{B'})-(N_c-2)]} z^x} H^U_{N_c-1, N_f}(t, \vec u, \vec v, z; R_{B'}).
\ee
Indeed, the Hilbert series of theory $B'$, dual to $B$, reads
\be
\begin{split}
	H^{(B)} &= H^{(B')} \\
	&=  (H_I^{(B')}+ H_{II}^{(B')})(t, \vec u, \vec v, z; R_{B'})  \\
	& =H^U_{N_c, N_f}(t, \vec u, \vec v, z; R_{B'}) + \frac{t^{[N_f(1-R_{B'})-(N_c-1)]} z^x}{1-t^{[N_f(1-R_{B'})-	(N_c-2)]} z^x}H^U_{N_c-1, N_f}(t, \vec u, \vec v, z; R_{B'}) \\
	&=H^{(A)}~.
\end{split}
\ee
Note that this expression matches with $H^{(A)}$ since $R_{B'}=R_{B}$.

\subsubsection{The Giveon--Kutasov duality}
In this subsection, we consider the following thories \cite{Giveon:2008zn}:
\paragraph{Theory $A$:} $U(N_c)_k$ gauge theory with $N_f$ flavours and $W=0$.
\paragraph{Theory $B$:} $U(N_f+|k|-N_c)_{-k}$ gauge theory with $N_f$ flavours $q$ and $\tilde{q}$, $N_f^2$ singlets $M$, and superpotential $W= M \tilde{q} q$.
\\~\\
For simplicity, let us assume that $k>0$ and that $N_f+|k|-N_c \geq 0$.  As pointed out in \cite[sec. 3.3]{Cremonesi:2016nbo}, the Coulomb branch is completely lifted and the full Hilbert series of theory $A$ is the mesonic Hilbert series of $U(N_c)$ with $N_f$ flavours:
\be \label{mesUNcNf}
\begin{split}
H^{(A)} &= H^U_{N_c,N_f} (t,\vec{u},\vec{v},y; r ) \\
&= \sum_{n_1, \ldots, n_{N_c} \geq 0}[0^{N_f-N_c-1},n_{N_c},\ldots,n_1; n_1, \ldots, n_{N_c}, 0^{N_f-N_c-1}]_{\vec{u},\vec{v}} (t^r y)^{2\sum_{j=1}^{N_c} {j n_j}} ~.
\end{split}
\ee 
This Hilbert series corresponds to the space generated by the mesons, which can be regarded as an $N_f \times N_f$ matrix, subject to the condition that the rank is at most $N_c$.

In theory $B$, the Coulomb branch is also lifted.  The moduli space is generated by $N_f \times N_f$ matrix $M$.  We shall argue that there is a quantum condition on the rank of $M$: $\mathrm{rank}(M) \leq N_c$ (this is the classical condition of the meson in theory $A$).  This can be seen as follows: If we give a VEV to $M$ of rank greater than $N_c$, say $N_c+p$ with $p>0$, then the lower energy theory is $U(N_f+|k|-N_c)_{-k}$ gauge theory with $N_f-N_c-p$ flavours; this can be described by the effective superpotential\footnote{Note that $q$ and $\tilde{q}$ have $R$-charge $1-r$, where $r$ is the $R$-charge for the electric quarks.  Hence, $\det q \tilde{q}$ has $R$-charge $2(1-r)(N_f-N_c-p)$.  Also, $V^\pm$ have $R$-charge $(N_f-N_c-p)r - (N_f+|k|-N_c-1)$.  Indeed, $(V_+ V_-\det q \tilde{q})^{\frac{1}{-p-|k|+1}}$ has $R$-charge 2.} $W \sim (V^+ V^-\det q \tilde{q})^{\frac{1}{-p-|k|+1}}$, where $V^\pm$ is the basic monopole operators in this low energy effective theory.  Since $p>0$ and $|k|>0$, we have runaway vacua.  We just matched the moduli space of theory $B$ with that of theory $A$.  Thus, the Hilbert series of theory $B$ is also given by \eref{mesUNcNf}.

\subsubsection{The Benini-Closset-Cremonesi (BCC) $[\vec{p},\mathbf{0}]a$ duality}
We consider the following theories \cite{Benini:2011mf}:
\paragraph{Theory $A$:} $U(N_c)_k$ gauge theory with $N_f$ fundamentals and $N_a$ antifundamentals such that $N_f >N_a$, $k=-\frac{1}{2}(N_f-N_a)$ and $W=0$.
\paragraph{Theory $B$:} $U(N_f-N_c)_{-k}$ gauge theory with $N_a$ fundamentals, $N_f$ antifundamentals, $N_f N_a$ singlets $M$, a singlet $S$, and superpotential $W= M \tilde{q} q +S \hat{X}^+$, where $\hat{X}^+$ is a basic monopole operator in theory $B$ with topological charge $+1$.
 \\~\\
  In theory $A$, the $U(1)$ gauge charges of the basic monopole operators $X^\pm$ are
 \be
\mp \left[ k \pm \frac{1}{2}(N_f - N_a) \right]  = \begin{cases} 0  &\qquad \text{for $X^+$} \\ -(N_f-N_a) & \qquad \text{for $X^-$} \end{cases}~.
 \ee
 Hence, the Coulomb branch that is generated by $X^-$ is lifted.  In theory $B$, the $F$-term $\partial_S W=0$ implies that $\hat{X}^+$ vanishes in the chiral ring.  We propose the following duality map:
\be
\begin{split}
\text{Theory $A$} &\qquad\quad \text{Theory $B$} \\
X^+ &\quad \longleftrightarrow \quad S~, \\
\text{mesons} &\quad \longleftrightarrow \quad M~.
\end{split}
\ee
 
Let us discuss about the $R$-charges of various fields.  The $R$-charge of the basic monopole operators in theory $A$ is
\be
R[X^\pm] = \frac{1}{2}N_f(1-r)+\frac{1}{2}N_a(1-r)-(N_c-1),
\ee
where $r$ is the $R$-charge of the quarks in theory $A$.  Since $X^+$ is mapped to $S$ under the duality, we have
\be
R[S]= R[X^\pm] = \frac{1}{2}N_f(1-r)+\frac{1}{2}N_a(1-r)-(N_c-1)~.
\ee
Since the mesons of theory $A$ are mapped to $M$ in theory $B$, we have 
\be
R[q] = R[\tilde{q}] =1-r~,
\ee
and so the monopole $V^+$ of theory $B$ has $R$-charge
\be
R[V^+]= {\blue \frac{1}{2}(N_f-N_a)} + \frac{1}{2}N_f r+\frac{1}{2}N_a r-(N_f -N_c-1)~,
\ee
where the blue term comes from the mixed gauge-$R$ symmetry CS terms; see \cite[(4.4)]{Amariti:2014lla}. Indeed,
\be
R[S]+R[\hat{X}^+] =2~;
\ee 
this is compatible with the superpotential term $S \hat{X}^+$ in theory $B$.

\subsection*{Theory $A$}
Since the Coulomb branch that is generated by $X^-$ is lifted in theory $A$, the monopole flux thus takes the form $(m_1, 0, \ldots, 0)$ with $m_1 \geq 0$.  
 \bi
\item If $m_1=0$, the residual theory is $U(N_c)$ gauge theory with $N_f$ fundamentals and $N_a$ antifundamentals, whose mesonic Hilbert series is
\be \label{mesUNcNfNa}
\begin{split}
& H^{\text{mes}}_{N_c,N_f,N_a} (t,\vec{u},\vec{v},y; r ) \\
&= \sum_{n_1, \ldots, n_{N_c} \geq 0}[0^{N_f-N_c},n_{N_c},\ldots,n_1; n_1, \ldots, n_{N_c}, 0^{N_a-N_c}]_{\vec{u},\vec{v}} (t^r y)^{\sum_{j=1}^{N_c} {j n_j}} ~.
\end{split}
\ee 
The mesonic chiral ring is generated by the $N_a × N_f$ meson matrix $M^{\tilde{a}}_{a} = \tilde{Q}^{\tilde{a}}_i Q^i_a$ of rank at most $N_c$.
\item If $m_1\neq 0$, the residual theory is $U(N_c-1)$ gauge theory with $N_f$ fundamentals and $N_a$ antifundamentals.  The mesons in this theory is to be dressed with the monopole operators generated by $X^+$.  The Hilbert series in this case is 
\be
H^{\text{mes}}_{N_c-1,N_f,N_a} (t,\vec{u},\vec{v},y; r ) \sum_{m_1=1}^\infty (a_{+})^{m_1}  = \frac{a_+}{1-a_+} H^{\text{mes}}_{N_c-1,N_f,N_a} (t,\vec{u},\vec{v},y; r )~,
\ee
where 
\be
a_+ = z t^{\frac{1}{2}(N_f+N_a)(1-r)-(N_c-1)} y^{-k_{gA}-\frac{1}{2}(N_f+N_a)}~.
\ee 
Here $z$ is the fugacity for the topological symmetry, $y$ is the fugacity for the axial symmetry, and $k_{gA}$ is the mixed Chern-Simons level between the central gauge $U(1)$ and the axial $U(1)_A$ symmetry, which is quantized to ensure that the exponent of $y$ is an integer.
\ei
The full Hilbert series of theory $A$ is therefore
\be
H^{(A)} = H^{\text{mes}}_{N_c,N_f,N_a} (t,\vec{u},\vec{v},y; r )+  \frac{a_+}{1-a_+} H^{\text{mes}}_{N_c-1,N_f,N_a} (t,\vec{u},\vec{v},y; r )~,
\ee
The chiral ring of the theory is generated by the $N_a \times N_f$ meson matrix $M$, of rank at most $N_c$, and by the bare monopole operators $X^+$ subject to the extra relation that the rank of $X^+ M$ is at most $N_c -1$, \ie~ $X^+ \mathrm{minor}_{N_c}(M) =0$.

\subsection*{Theory $B$}
Now let us consider theory $B$.  The $U(1)$ gauge charges of the basic monopole operators $V^\pm$ are
 \be
\mp \left[ -k \pm \frac{1}{2}(N_a - N_f) \right]  = \begin{cases} 0  &\qquad \text{for $\hat{X}^+$} \\ N_f-N_a & \qquad \text{for $\hat{X}^-$} \end{cases}~.
 \ee
Hence $\hat{X}^-$ is not in the chiral ring and the Coulomb branch parametrised by $\hat{X}^-$ is lifted.  There are quantum conditions that give bounds on the rank of $M$.  These correspond to the classical conditions for the rank of the mesons in theory $A$.  We can derive such quantum conditions below.  

Since the $F$-terms with respect to $S$ implies that $\hat{X}^+$ vanishes in the chiral ring.  The moduli space of theory $B$ is generated by the $M$.  Therefore, after imposing such quantum conditions on $M$, we conclude that the Hilbert series of theory $B$ is equal to that of theory $A$. 

Let us turn on a VEV of $M$ with rank $N_c+p$.  The low energy theory is described by $U(N_f-N_c)_{-k}$ gauge theory with $N_a-N_c-p$ fundamentals and $N_f-N_c-p$ antifundamentals. We can use the topological and axial symmetries to constraint the form of the effective superpotential.  We claim that the only possible consistent combination that can appear in the effective superpotential is
\be
W'_B = (S \hat{X}^+)^{P}
\ee
for some power $P$, which can be worked out from the $R$-charges of $S$ and $\hat{X}^+$ as follows.  Since the singlets $M$ are mapped to the mesons in theory $A$, for $p < 0$, the gauge group $U(N_c)$ of theory $A$ is broken to $U(|p|)$. On the other hand, for $p\geq 0$, $U(N_c)$ is completely broken.  Since $S$ is mapped to the monopole operator $X^+$, the $R$-charges of $S$ is given as follows:
\be
R[S] = \begin{cases} 
\frac{1}{2}(N_a-N_c-p)r + \frac{1}{2}(N_f-N_c-p)r -(-p-1)~, &\qquad ~p \leq -1 \\
\frac{1}{2}(N_a-N_c-p)r + \frac{1}{2}(N_f-N_c-p)r~,  &\qquad~p \geq 0
\end{cases}
\ee
where the $R$-charges of the magnetic quarks are $1-r$.  The $R$-charges of $\hat{X}^+$ can be computed as usual
\be
\begin{split}
R[\hat{X}^+] &= {\blue \frac{1}{2}\left[ (N_f-N_c-p)- (N_a-N_c-p) \right]} + \frac{1}{2}(N_f-N_c-p) r \\
& \qquad +\frac{1}{2}(N_a-N_c-p)  r-(N_f -N_c-1)~.
\end{split}
\ee
For $p\leq -1$, we see that $R[S]+R[\hat{X}^+]= 2$.  For $p\geq 0$, we have $R[S]+R[\hat{X}^+] = 1-p$.  Hence, the power $P$ is $1$ for $p \leq -1$ and $\frac{2}{1-p}$ for $p\geq 0$:
\be
W'_B = 
\begin{cases}
S \hat{X}^+ ~ , &\qquad ~p \leq -1\\
(S \hat{X}^+)^{\frac{2}{1-p}} ~,  &\qquad~p \geq 0~.
\end{cases} 
\ee
For $p\geq 1$, we have runaway vacua.  This agrees with the analysis of the theory $A$, which says that the mesons have rank at most $N_c$.  

For $p\leq 0$, we have a positive power of $S \hat{X}^+$ in the superpotential.  If $\hat{X}^+ \neq 0$, we have $S=0$ and this corresponds to the case of $m_1=0$ in theory $A$.  If $\hat{X}^+ =0$, then the value of $S$ is arbitrary; when $S\neq 0$, this corresponds to the case of $m_1 \neq 0$.  This is in agreement with theory $A$.

\subsubsection{The duality involving $W = X^-$, chiral flavours and Chern--Simons terms} \label{sec:XmCS}
Let us consider the following theories \cite{Benini:2017dud}:
\paragraph{Theory $A$:} $U(N_c)_{\frac{k}{2}}$ with $k>0$ and $(N_f, N_a=N_f-k)$ fund/antifund and $W=X^-$. 
\paragraph{Theory $B$:} $U(N_f-N_c-1)_{-k/2}$ with $(N_f, N_a=N_f-k)$ fund/antifund, $N_fN_a$ singlets and
superpotential $W=\hat X^+ + \sum_{i}^{N_f}\sum_{j}^{N_a}{M^i}_j \tilde{q}_i q^j$.
\\~\\
In theory $A$ the $U(1)$ gauge charge of the monopoles are given by
\be
\mp \left[ \frac{k}{2} \pm \frac{1}{2}(N_f - N_a) \right]  = \begin{cases} -k  &\qquad \text{for $X^+$} \\ 0 & \qquad \text{for $X^-$} \end{cases}~.
\ee
Thus, the Coulomb branch generated by $X^+$ is lifted due to the non-zero Chern-Simons level.   However, by putting $X^-$ in the superpotential, we can write $X^-=e^{Z^-}$ and the $F$-term with respect to $Z^-$ implies that $X^-$ vanishes in the chiral ring.  Thus,  the Coulomb branch generated by $X^-$ is also lifted.   The residual theory is the full $U(N_c)$ gauge theory with $(N_f, N_a)$ fund/antifund flavours, whose mesonic Hilbert series is given by
\bea
H^{(A)}&=H^{\text{mes}}_{N_c, N_f, N_a}(t, \vec{u}, \vec{v}; R)= \notag \\
&\sum_{n_1, \dots, n_{N_c}\ge0} [0^{N_f-N_c}, n_{N_c}, \dots, n_1; n_1, \dots, n_{N_c}, 0^{N_a-N_c}]_{\vec u, \vec v}
(t^R)^{2 \sum_j jn_j}, \label{HSABBP}
\eea
where observe that there is no fugacity for the axial symmetry since it is broken by the presence of $X^-$ in the 
superpotential.  The $R$-charge $R$ of the quarks and antiquarks are fixed by the monopole superpotential:
\be
2 = \frac{1}{2} N_f (1-R) +\frac{1}{2} N_a (1-R) -(N_c-1) \quad \Rightarrow \quad R =1-2 \frac{N_c+1}{N_a+N_f}~.
\ee
The generators of the mesonic chiral ring are the mesons, which are matrix $N_f \times N_a$ of rank at most 
$N_c$.

In theory $B$ the Coulomb branch is lifted for the same reason of theory $A$ (\ie~ due to both the non-zero Chern-Simons level and the monopole superpotential). The moduli space is generated by the singlets $M$, which can be viewed as an $N_f \times N_a$ matrices.  We shall argue that there is a quantum condition on the rank of $M$: $\mathrm{rank}(M) \leq N_c$ (this is the classical condition of the meson in theory $A$).  Let us give a VEV to $M$ with rank $N_c+p$ with $p>0$. The low energy effective theory is a $U(N_f-N_c-1)_{-k/2}$ gauge theory with $(N_f'=N_f-N_c-p, \,\, N_a'=N_f-k-N_c-p)$ fund/antifund.   At this step, we can use the BCC duality to obtain the dual theory.  Since the dual gauge group is the unitary group of rank $N_f'-(N_f-N_c-1) = -p+1$, we see that for $p>1$, supersymmetry is broken.  For $p=1$, the dual theory is a WZ model with singlets $S$ and $\hat{X}^+$ with superpotential $W=S \hat{X}^+$.  In which case the effective superpotential for theory $B$ is
\be
W'_B = \hat X^++ S \hat{X}^+~,
\ee
where the first term of $W'_B$ comes from the first term of the original superpotential of theory $B$.  The equations of motion imply that both $S$ and $X^+$ are massive and we do not have a supersymmetric vacuum.

\subsubsection{The Aharony duality for symplectic gauge groups}
We consider the following duality, which was proposed in \cite{Aharony:1997gp}:
\paragraph{Theory $A$:} $USp(2N_c)$ with $2N_f$ fundamental chirals $Q_i$ and superpotential $W=0$. 
\paragraph{Theory $B$:} $USp(2(N_f-N_c-1))$ with $2N_f$ fundamental chirals $q_i$, $N_f(2N_f -1)$ singlets 
$M$ and singlet $Y$ (that are dual to the mesons and the monopole of theory $A$) and superpotential 
$W=M q q + \hat Y S$, where $\hat Y$ is the fundamental monopole of theory $B$.
\\~\\
Let us fist study $R$-charges of various fields. In theory $A$, let the $R$-charge of the fundamentals be $R[Q] = R$, so that the monopole operator have the $R[Y]=2N_f (1-R) -2N_c$.  Since the mesons in theory $A$ are mapped to $M$ in theory $B$, the superpotential in theory $B$ gives $R[q] \equiv 1-R$, and so
$R[\hat Y]=2N_f R -2 (N_f-N_c-1)$. Thus, by the superpotential it follows that the $R$-charge of the singlet $S$ is 
$R[S]=2-R[\hat Y]=2-2N_f R+2(N_f-N_c-1) = 2N_f(1-R)-2N_c = R[Y]$.   Indeed, this is consistent with the expectation that the monopole operator $Y$ in theory $A$ is mapped to the singlet $S$ in theory $B$ under the duality.

\subsubsection*{Theory $A$}

In the following we discuss the Hilbert series of theory $A$.  This has been analysed in detail in Section 6 of \cite{Cremonesi:2015dja}.  It has two contributions, depending on the value of the magnetic charge $m$ in the magnetic flux $(m,0^{N_c-1})$:
\begin{itemize}
	\item{$m=0$: the residual theory is $USp(2N_c)$ with $2N_f$ fundamentals, whose mesonic Hilbert series reads \cite{Hanany:2008kn}:
	\be \label{HUSp}
	\begin{split}
		& H^{I(A)}(t, y, \vec x; R) \\
		&=H^{USp}_{2N_c, 2N_f}(t, y, \vec x; R)  \\
		&=\sum_{n_2, n_4, \dots, n_{2N_c}}[0, n_2, 0, n_4, \dots, 0, n_{2N_c}, 0^{2(N_f-N_c)-1}]_{\vec x} 
	(t^R y)^{2\sum_{j=1}^{N_c} jn_{2j}} ~. 
	\end{split}
	\ee}
	\item{$m>0:$ the residual theory is $USp(2(N_c-1))$ with $2N_f$ fundamentals; since now the magnetic flux
	is non vanishing the Hilbert series contains a dressing factor taking into account of the monopole $Y^m$ with flux $(m,0^{N_c-1})$ such that $m\geq1$:
	\bea
		H^{II(A)}(t, y, \vec x; R)&=\left(\sum_{m=1}^{\infty} t^{R[Y^m]} y^{A[Y^m]}\right) H^{USp}_{2(N_c-1), 2N_f}(t,y,\vec x; R) 		\notag \\
		&=\left(\sum_{m=1}^{\infty} t^{[2N_f(1-R)-2N_c]m}y^{-2N_f m} \right)H^{USp}_{2(N_c-1), 2N_f}(t,y, \vec x; R)\notag \\
		&=\frac{t^{[2N_f(1-R)-2N_c]} y^{-2N_f}}{1-t^{[2N_f(1-R)-2N_c]} y^{-2N_f}}H^{USp}_{2(N_c-1), 2N_f}(t,y, \vec x; R),
	\eea
	where $y$ is the fugacity for the $U(1)$ axial symmetry.
	}
\end{itemize}
Thus the total Hilbert series of the theory is the given by adding the two contributions $H^{I(A)}$ and $H^{II(A)}$:
\bea
	H^{(A)}(t, y, \vec x; R)&=(H^{I(A)}+H^{II(A)})(t, y, \vec x; R)\notag \\
	&=H^{USp}_{2N_c, 2N_f}(t, y, \vec x; R)+
	\frac{t^{[2N_f(1-R)-2N_c]} y^{-2N_f}}{1-t^{[2N_f(1-R)-2N_c]} y^{-2N_f}}H^{USp}_{2N_c, 2N_f}(t,y, \vec x; R)
\eea
The Hilbert series tells us that the moduli space is generated by the antisymmetric $2N_c \times 2N_c$ meson matrix $M$ and by the fundamental monopole operator $Y$, subject to the condition:
\be
\begin{split}
\epsilon_{i_1 \cdots i_{2N_f}} M^{i_1 i_2} \cdots M^{i_{2N_c+1} i_{2N_c+2}} = 0~, \\
Y \epsilon_{i_1 \cdots i_{2N_f}} M^{i_1 i_2} \cdots M^{i_{2N_c-1} i_{2N_c}} = 0~.
\end{split}
\ee
Note that the first equality implies that the rank of $M$ is at most $2N_c$.  The second equality implies that for $Y\neq 0$, the rank of $M$ is at most $2(N_c-1)$.

\subsubsection*{Theory $B$}
Let us now analyse the moduli space of theory $B$.  Since the mesons are mapped to the singlets $M$, we give a VEV to $M$ of rank $2(N_c+p)$ and study at the moduli space for various values of $p$. The low energy effective theory is  a $USp(2(N_f-N_c-1))$ gauge theory with $2(N_f-N_c-p)$ massless quarks\footnote{In this theory, the $R$-charges of the monopole operator $\hat{Y}$ and $\text{Pf}(qq)$ are
\bea
	&R[\hat Y]=2(N_f-N_c-p)R - (N_f-N_c-1)~, \nn 
	&R[\text{Pf}(qq)]=2(1-R)(N_f-N_c-p)~, \nn
\eea
and so the combination $\hat Y \text{Pf}(qq)$ has $R-$charge $2(1-p)$.}.  Below we analyse the possible cases of $p$.  We shall see that, for $p>0$, there is no stable supersymmetric vacua, whereas for $p \leq 0$, the moduli space agrees with that of theory $A$.
\begin{itemize}
	\item{$p<0$:  The effective description away from the origin of the moduli space is given by the superpotential
	\be\label{Wplesszero}
		W=\hat Y S + (\hat Y \text{Pf}(qq))^{\tfrac{1}{1-p}}.
	\ee
	The equations of motion are the following
	\bea
		\label{FY}&\d_Y W=0: \quad \hat Y=0, \\
		&\d_{\hat Y} W=0: \quad S+\frac{1}{1-p}\text{Pf}(qq)^{\tfrac{1}{1-p}} 
		\hat Y^{\tfrac{p}{1-p}}=0,
	\eea
	Equation (\ref{FY}) implies that the Coulomb branch of theory $B$ is completely lifted. Since the superpotential
	(\ref{Wplesszero}) is valid away from the origin, $S$ can take any arbitrary VEV. According to this observation 
	we can have the following two cases: The case of $S=0$ corresponds to the magnetic flux $m=0$ of 
	the electric theory, where the corresponding Hilbert series is $H^{I(A)}$, and the case of $S \ne 0$ corresponds to $m>0$, where the Hilbert series is equal to $H^{II(A)}$.}
	\item{$p=0$: the effective superpotential of the form:
	\be
		W=\hat Y S - \hat Y \text{Pf}(qq).
	\ee
	The equations of motion for $S$ and $\hat Y$ are 
	\bea
		&\d_S W=0: \quad \hat Y=0, \\
		&\d_{\hat Y} W=0: \quad \hat Y(S-\text{Pf}(qq))=0,
	\eea
	whose solution is $S=\hat Y=\text{Pf}(qq)=0$. Recalling that the singlet $S$ is mapped to the basic monopole operator $Y$ of theory $A$, this case correspond to the magnetic flux $m=0$ in theory $A$, and hence the corresponding Hilbert series is $H^{I(A)}$.}
	\item{$p=1$: In this case instantons in the low energy theory generate the constraint $\hat Y \text{Pf}(qq)=1$,
	which can be put in the superpotential through a Lagrange multiplier $\lambda$:
	\be
		W=\hat Y S + \lambda (\hat Y \text{Pf}(qq)-1).
	\ee
	Using the solution for $\partial_\lambda W=0$, we can rewrite the superpotential as follows
	\be
		W=\frac{S}{\text{Pf}(qq)},
	\ee 
	so we have runway vacua.
	}
	\item{$p>1$: The effective superpotential is given by
	\be
		W=\hat Y S + (\hat Y \text{Pf}(qq))^{\tfrac{1}{1-p}}.
	\ee
	The equation of motion for the basic monopole reads
	\be
		\d_{\hat Y} W=0: \quad \hat Y=[(p-1) S]^{\tfrac{1-p}{p}} \text{Pf}(qq)^{-\tfrac{1}{p}}.
	\ee
	Substituting into the superpotential we get
	\be
		W \sim \left(\frac{S}{\text{Pf}(qq)}\right)^{\tfrac{1}{p}},
	\ee
	and, again, we have runway vacua.}
\end{itemize}

\subsubsection{The duality involving symplectic gauge groups and $W=Y$}
We consider the following duality \cite{Aharony:2013dha}:
\paragraph{Theory $A$:} $USp(2N_c)$ with $2N_f$ fundamentals and superpotential $W=Y$.
\paragraph{Theory $B$:} $USp(2(N_f-N_c-2))$ with $2N_f$ fundamentals, $N_f(2N_f-1)$ singlets $M$, and superpotential $W=M qq +\hat Y$. 
\\~\\
In theory $A$, { due to the presence of $Y$ in the superpotential, the Coulomb branch is lifted and $Y=0$ in the chiral ring}.  The Hilbert series thus get the mesonic contribution:
\be
	H^{(A)}=H^{USp}_{2N_c, 2N_f}(t, y, \vec x; R),
\ee
where $R$ is the $R$-charge of the quarks.
This means that the moduli space is generated by the mesons $M$, with the constraint $\text{rank}(M)\le 2N_c$. The $R$-charge of the meson 
is fixed the presence of the monopole in the superpotential as usual, since $R[Y]=2$ and is also given in terms of the 
$R$-charge of the quarks:
\be	
	R[Y]=2N_f(1-R)-2N_c=2,
\ee
from which we get 
\be \label{USpelec}
	R[M]=2R=2\frac{N_f-N_c-1}{N_f}.
\ee

In theory $B$, the $R$-charge $r$ of the fundamentals is fixed by the monopole superpotential term $\hat Y$, whose $R$-charge is $2$:
\be
	R[\hat Y]=2=2N_f(1-r)-2(N_f-N_c-2)~.
\ee
Therefore, the $R$-charge of the singlets $M$ is
\be
	R[M]=2-2r=2\frac{N_f-N_c-1}{N_f}~,
\ee
in agreement with \eref{USpelec}.

In order to analyse the moduli space of theory $B$, let us now perform the Aharony duality.  We get a $USp(2(N_c+1))$ gauge theory with $2N_f$ fundamentals, singlet $S$ and superpotential $W'=S\hat Y + S$, where $\hat Y$ are the basic monopoles of this theory. The $F-$terms $\d_{S}W'=0$ gives a non-zero VEV to the monopole $\hat Y=-1$.  Since the vacuum expectation value of $\hat Y$ is non-zero, the gauge group $USp(2(N_c+1))$  is broken to $USp(2N_c)$, with an additional $U(1)$ factor which decouples in the IR. If we substitute back $\hat Y=-1$ to the superpotential, we end up with $W=0$.   Thus, the residual theory is a $USp(2N_c)$ gauge theory with $N_f$ flavours, whose mesonic Hilbert series is given by $H^{USp}_{2N_c, 2N_f}(t, y, \vec x; R)$.  This is indeed in agreement with the Hilbert series of theory $A$.

\subsubsection{The BCC duality for orthogonal gauge groups}
Let us consider the following duality \cite{Benini:2011mf, Aharony:2011ci}: 
\paragraph{Theory $A$:} $O(N_c)$ with $N_f$ chirals in the vector representation and zero superpotential.
\paragraph{Theory $B$:} $O(N_f-N_c+2)$ with $N_f$ chirals in the vector representation, $N_f(2N_f+1)$ singlets $M$,  a singlet $S$ and superpotential $W=M qq +S \hat Y$. 
\\~\\
The $R$-charge of the monopole operator in theories with orthogonal gauge group reads
\be
	R[Y]=N_f(1-R)-(N_c-2).
\ee
\subsubsection*{Theory $A$}

The Coulomb branch is parametrized by the fundamental monopole operator $Y$. The magnetic flux $(m, 0^{n-1})$ (where $n$ is given by $N_c=2n$ for the even case and $N_c=2n+1$ for the odd case) of $Y^m$ gives the following two contributions to the Hilbert series
\begin{itemize}
	\item{$m=0$: the residual theory is $O(N_c)$ gauge theory with $N_f$ flavours and, in this case, the Hilbert series is
	\be
		H^{I(A)}(t, y, \vec x; R)=H^{O}_{N_c, N_f}(t, y, \vec x; R)~,
	\ee
	where $H^{O}_{N_c, N_f}(t, y, \vec x; R)$ is the mesonic Hilbert series of the aforementioned residual theory.  It can be obtained from that of the $SO(N_c)$ gauge theory with $N_f$ flavours \cite[(2.29)]{Hanany:2008kn} by projecting out the baryons and reads:
	\bea \label{mesO}
		&H^{O}_{N_c, N_f}(t, y, \vec x; R)= \\ \notag 
		&=\sum_{n_1, n_2, \dots, n_{N_c}\ge 0}[2n_1, 2n_2, \dots, 2 n_{N_c}, 0^{N_f-N_c-1}]_{\vec x} 
		(t^{R} y)^{2\sum_{j=1}^{N_c} jn_{j}}~,
	\eea  
	where $y$ is the fugacity for the axial symmetry and $\vec{x}$ denotes the fugacities associated with the flavour symmetry $SU(N_f)$.
	}
	\item{$m>0$: the residual theory is $O(N_c-2)$ gauge theory\footnote{The gauge group $O(N_c)$ is actually broken to $O(N_c-2) \times O(2)$, with $O(2)$ decoupled.} with $N_f$ flavours and, in this case, the Hilbert series is the mesonic contains a dressing factor taking into account of the monopole 
	operators $Y^m$:
	\bea
		H^{II(A)}(t, y, \vec x; R)&=\left(\sum_{m=1}^\infty t^{R[Y^m]} y^{A[Y^m]} \right) 
		H^{O}_{N_c-2, N_f} (t, y, \vec x; R)\notag \\
		&=\left(\sum_{m=1}^\infty t^{(N_f-N_c+2-N_f R)m} y^{-N_f m} \right) H^{O}_{N_c-2, N_f}(t, y, \vec x; R) \notag 		\\
		&=\frac{t^{(N_f-N_c+2-N_f R)} y^{-N_f}}{1-t^{(N_f-N_c+2-N_f R)} y^{-N_f}}H^{O}_{N_c-2, N_f}(t, y, \vec x; R) 		\notag \\
	\eea}
\end{itemize}
Hence the Hilbert series of the theory reads
\bea \label{totHSAO}
	H^{(A)}(t, y, \vec x; R)&=(H^{I(A)}+H^{II(A)})(t, y, \vec x; R)\notag \\
	&=H^{O}_{N_c, N_f}(t, y, \vec x; R)+\frac{t^{(N_f-N_c+2-N_f R)} y^{-N_f}}{1-t^{(N_f-N_c+2-N_f 	R)} y^{-N_f}}H^{O}	_{N_c-2, N_f}(t, y, \vec x; R).
\eea
The Hilbert series tells us that the moduli space of theory $A$ is generated by the symmetric $N_f \times N_f$ meson matrix $M$ and by the fundamental monopole $Y$, subject to the following  relations
\be
\begin{split}
	\epsilon^{i_1 i_2 \cdots i_{N_f}} \epsilon^{j_1 j_2 \cdots j_{N_f}}  M_{i_1 j_1} \cdots M_{i_{N_c+1} j_{N_c+1}}&=0, \\
	Y \epsilon^{i_1 i_2 \cdots i_{N_f}} \epsilon^{j_1 j_2 \cdots j_{N_f}}  M_{i_1 j_1} \cdots M_{i_{N_c-1} j_{N_c-1}} &=0~.
\end{split}
\ee
The first relation implies that the mesons have at most rank $N_c$, while the second implies that, for $Y \ne 0$,
$M$ has at most rank $N_c-2$. 

\subsubsection*{Theory $B$}
Since the mesons are mapped to the singlets $M$ we give a VEV to $M$ of rank $N_c+p$ and we study the moduli space for various $p$. The low energy effective theory is $O(N_f-N_c+2)$ with $N_f-N_c-p$ massless quarks\footnote{The $R-$charges of $\hat Y$ and $\text{det}(qq)$ are the given by:
\be
	R[\hat Y]=(N_f-N_c-p)(1-r)-[(N_f-N_c+2)-2], \qquad R[\text{det}(qq)]=2r(N_f-N_c-p)~, \nn
\ee
where $R[q]=r$.  The combination $\hat Y^2 \text{det}(qq)$ thus has $R$-charge $-2p$.}.
In the following analysis of the moduli space we shall see that for $p\ge0$ there is no stable supersymmetric vacua, while for $p<0$ the moduli space agrees with the one in theory $A$. Let us analyze the various cases we get depending on $p$: 
\begin{itemize}
	\item{$p<0$: the $F$-terms we get from the effective superpotential $W=S\hat Y + \left(\hat Y^2\text{det}(qq)		\right)^{-\tfrac{1}{p}}$ are 
	\bea
		&\d_S W=0: \quad \hat Y=0, \\
		&\d_{\hat Y} W=0: \quad S=\frac{2}{p}(\text{det}(qq))^{-\tfrac{1}{p}} \hat Y^{-\tfrac{p+2}{p}}.
	\eea
	The first of these equations implies that the Coulomb branch is completely lifted, and, since the effective 
	superpotential is valid away from the origin of the moduli space the singlet $S$ can take 
	any arbitrary VEV. We can have two cases: for $S=0$ we recover the case of magnetic flux $m=0$ in electric 		theory, corresponding with the Hilbert series $H^{I(A)}$, while $S\ne0$ corresponds $m>0$, where the Hilbert 
	series is equal to $H^{II(A)}$.}
	\item{$p=0$: the $R$-charges of $\hat Y$ and $\text{det}(qq)$ implies that the combination 
	$\hat Y^2 \text{det}(qq)$ has zero $R$-charge, meaning that there is  
	a constraint $\hat Y^2 \text{det}(qq)=1$ generated.  Thus, we have the effective superpotential 
	\be
		W=S \hat Y + \lambda (\hat Y^2 \text{det}(qq)-1).
	\ee
	where $\lambda$ is the Lagrange multiplier.  Using the equation of motion, we have
	\be
		W=\pm \frac{S}{{\text{det}(qq)}^{1/2}},
	\ee
	which gives runway vacua. 
	} 
	\item{$p\ge1$: the effective superpotential reads
	\be
		W=S\hat Y + \left(\hat Y^2\text{det}(qq)\right)^{-\tfrac{1}{p}}.
	\ee
	The equations of motion for $\hat{Y}$ reads
	\be
		\hat Y=\left[ \frac{p}{2} \text{det}(qq)^{1/p} S \right]^{-\tfrac{p}{p+2}}.
	\ee
	Substituting into the superpotential we finally obtain 
	\be
		W \sim \left(\frac{S^2}{ \text{det}(qq)}\right)^{\tfrac{1}{2+p}},
	\ee
	which again gives runway vacua.}
\end{itemize}

\subsubsection{The ARSW duality for special orthogonal gauge groups}
In this section we consider the following duality:
\paragraph{Theory $A$:} $SO(N_c)$ with $N_f$ chirals in the vector representation and zero superpotential.
\paragraph{Theory $B$:} $SO(N_f-N_c+2)$ with $N_f$ chirals in the vector representation, $N_f(2N_f+1)$ singlets $M$,  a singlet $Y$ and superpotential $W=M qq +Y \hat Y$, where $\hat{Y}$ is the basic monopole operator in this theory.
\\~\\
This duality was proposed by Aharony, Razamat, Seiberg and Willet (ARSW) in \cite{Aharony:2013kma}.  A crucial difference between this duality and the duality involving gauge group $O(N_c)$, discussed in the previous section, is the presence of the baryons and the baryon monopoles in the former.

\subsubsection*{Theory $A$}
Let us start with theory $A$. The magnetic flux of the monopole operator $Y$ takes the form $(m,0,\ldots,0)$ in the Dynkin label notation.  

For $m= 0$, the residual theory is $SO(N_c)$ with $N_f$ flavours (for $N_f \geq N_c$), in which case the Hilbert series is given by \cite[(2.29)]{Hanany:2008kn}:
\bea \label{HSO}
&H^{SO}_{N_c, N_f}(t, y, \vec x; R) \nn \\
&=  
\sum_{n_1, n_2, \dots, n_{N_c}\ge 0}[2n_1, 2n_2, \dots, 2n_{N_c-1}, n_{N_c}, 0^{N_f-N_c-1}]_{\vec x} 
		(t^{R} y)^{2\sum_{j=1}^{N_c-1} jn_{j} + n_{N_c} N_c}~, 
\eea
where $R$ is the $R$-charge for the quarks and $y$ is the fugacity for the axial symmetry.  We emphasise that $H^{SO}_{N_c, N_f}(t, y, \vec x; R)$ counts the operators generated by the mesons and the baryons, subject to algebraic relations among themselves.

For $m \neq 0$, the gauge group is broken to $S(O(N_c-2) \times O(2))$; this includes a $\BZ_2$ group corresponding to transformations with determinant $(-1)$ both in $SO(N_c -2)$ and in $SO(2)$.  As discussed in \cite{Aharony:2013kma}, the gauge invariant monopole operator Y is charge conjugation even in $SO(2)$, and it will be denoted by $W_+$.  There is also a charge conjugation odd monopole operator $W_-$ in $SO(2)$.  In order to obtain an invariant quantity under the $\BZ_2$ part of the gauge group, one can form a baryon monopole $\beta = Q^{N_c-2} W_-$, where the product $Q^{N_c-2}$ is invariant under the $SO(N_c-2) \times SO(2)$ residual gauge symmetry.  The $R$-charges and the $U(1)_A$ charges of the monopole operators are as follows:
\be
\begin{split}
R[Y]&= R[W^\pm] = (N_f-(N_c-2))(1-R)~, \\
R[\beta] & = (N_f-N_c+2)-(N_f-2N_c+4)R~, \\
A[Y]&= A[W^\pm] = -N_f~,\\
A[\beta] &= N_c-N_f-2~,
\end{split}
\ee
The Hilbert series for $m\neq 0$ can be obtained as follows:
\be
\begin{split}
& \left(\sum_{m=1}^\infty t^{m R[W_+]} y^{m A[W_+]} \right) H^{SO}_{N_c-2, N_f}(t, y, \vec x; R)  \\
&= \left( \frac{t^{(N_f-N_c+2)(1-R)} y^{-N_f}}{1-t^{(N_f-N_c+2)(1-R)} y^{-N_f} }  \right) H^{SO}_{N_c-2, N_f}(t, y, \vec x; R) ~,
\end{split}
\ee
where the factor in the bracket is the dressing factor coming from the monopole operator $Y^m = W_+^m$ with $m\geq 1$.


The total Hilbert series of theory $A$ can be obtained in a similar way as in \eref{totHSAO}:
\be \label{totalSOA}
H^{(A)}(t,y, \vec x; R) = H^{SO}_{N_c, N_f}(t, y, \vec x; R) + \left( \frac{t^{(N_f-N_c+2)(1-R)} y^{-N_f}}{1-t^{(N_f-N_c+2)(1-R)} y^{-N_f} }  \right) H^{SO}_{N_c-2, N_f}(t, y, \vec x; R)~,
\ee
Note that if the charge conjugation symmetry is gauged, we recover formula \eref{totHSAO} for the $O(N_c)$ gauge group. 

\paragraph{The special case of $N_f=N_c-2$.}  In this case we obtain
\be 
\begin{split}
&H^{SO}_{N_c-2, N_c-2}(t, y, \vec x; R) \\
&= \sum_{n_1, n_2, \dots, n_{N_c-2}\ge 0}[2n_1, \dots, 2n_{N_c-3}]_{\vec x} 
		(t^{R} y)^{2\sum_{j=1}^{N_c-3} jn_{j} + n_{N_c-2} (N_c-2)} \\
&= \frac{1-(t^R y)^{2(N_c-2)} }{1-(t^R y)^{(N_c-2)} } H^O_{N_c-2, N_c-2} (t, y, \vec x; R) ~,
\end{split}
\ee
where $H^O_{N_c-2, N_c-2} (t, y, \vec x; R)$ can be computed using \eref{mesO} to obtain
\be
H^O_{N_c-2, N_c-2} (t, y, \vec x; R)  = \PE \left[[2,0,\ldots,0] t^{2R} y^2 \right]~,
\ee
which is a generating function of the mesons $M$.  The factor $(1-(t^R y)^{(N_c-2)})^{-1}$ corresponds to the baryon $B=Q^{N_c-2}$, and the numerator $1-(t^R y)^{2(N_c-2)}$ indicates that there is a chiral ring relation $B^2= \det_{(N_c-2) \times (N_c-2)}(M)$. The total Hilbert series \eref{totalSOA} in this case is therefore
\be \label{totHSNcm2}
\begin{split}
&H^O_{N_c-2, N_c-2} (t, y, \vec x; R)+ \frac{y^{-(N_c-2)}}{1-y^{-(N_c-2)}} \frac{1-(t^R y)^{2(N_c-2)} }{1-(t^R y)^{(N_c-2)} } H^O_{N_c-2, N_c-2} (t, y, \vec x; R) \\
& = \frac{1-t^{2(N_c-2)R}}{(1-y^{-(N_c-2)} )( 1- t^{(N_c-2)R})} H^O_{N_c-2, N_c-2} (t, y, \vec x; R)~. 
\end{split}
\ee
where we have used the fact that there is no baryon in $SO(N_c)$ gauge theory with $N_c-2$ flavours and so we have the following equality \cite[(2.28)]{Hanany:2008kn}:
\be
H^{SO}_{N_c, N_c-2} (t, y, \vec x; R) = H^O_{N_c-2, N_c-2} (t, y, \vec x; R)~.
\ee
Let us discuss the physical interpretation of \eref{totHSNcm2}.  The moduli space of this theory is a complete intersection, generated by the monopole operator $Y$, the baryon monopole operator $\beta$, and the mesons $M$.  These generators correspond to the factors $\left(1-y^{-(N_c-2)} \right)^{-1}$, $\left(1-t^{(N_c-2)R} \right)^{-1}$ and $H^O_{N_c-2, N_c-2} (t, y, \vec x)$ in the Hilbert series, respectively.  The numerator $1-t^{2(N_c-2)R}$ indicates that there is a chiral ring relation 
\be
\beta^2 \sim Y^2 \det(M)~,
\ee
in agreement with \cite[(2.22)]{Aharony:2013kma}.  We emphasise that, in the Hilbert series \eref{totHSNcm2}, the factor corresponding to the baryon monopole $\beta$ emerges only after summing the second term with the first term.


\subsubsection*{Theory $B$}
The gauge invariant combination $qq$ in theory $B$ vanishes due to the $F$-term $\partial_M W=0$.  The singlets $M$ are subject to the quantum relation $\mathrm{rank}(M) \leq N_c$, which can be derived in the same way as in the previous subsection for orthogonal gauge groups.  Also, the singlet $Y$ in theory $B$ is mapped to the monopole operator $Y$ in theory $A$.  As discussed in \cite[(2.28)]{Aharony:2013kma}, the baryon monopole operators and the baryons in theory $B$ are mapped to the baryons and the baryon monopole operators in theory $A$, and the former satisfy the same set of relations as those in theory $A$.  As a consequence, the Hilbert series of theory $B$ is equal to that of theory $A$.

\subsubsection{The duality involving orthogonal gauge groups and $W=Y$}
In this section we consider the following duality \cite{Aharony:2013kma}:
\paragraph{Theory $A$:} $O(N_c)$ with $N_f$ chirals in the vector representation and superpotential $W=Y$.
\paragraph{Theory $B$:} $O(N_f-N_c)$ with $N_f$ chirals in the vector representation, $N_f(2N_f+1)$ singlets $M$ and superpotential $W=M qq + \hat Y$. 
\\~\\
As usual, the $R$-charge $R$ of the quarks in theory $A$ is fixed by the superpotential, which gives 
\be
	R[Y]= 2 = N_f(1-R) -(N_c-2) \qquad \Rightarrow \qquad R =\frac{N_f-N_c}{N_f}.
\ee
{ Due to the presence of $Y$ in the superpotential, the Coulomb branch is lifted and we set $Y=0$ in the chiral ring.}  Hence, the Hilbert series of theory $A$ is given by the mesonic Hilbert series of $O(N_c)$ gauge theory with $N_f$ flavours
\be
	H^{(A)}(t, y, \vec x; R)=H^O_{N_c, N_f}(t, y, \vec x; R)~,
\ee
where $H^O_{N_c, N_f}(t, y, \vec x; R)$ is given by \eref{mesO}.

Let us now analyse theory $B$.  It is convenient to consider its BCC dual, which is an $O(N_c+2)$ theory with $N_f$ fundamentals and a singlet $\hat Y$ and with a superpotential $W=\hat Y Y'+ \hat Y$, where $Y'$ is now the basic monopole operator of this new theory.
Taking the equations of motion of the singlet we see that the VEV of the monopole is fixed, $Y'=-1$, hence the 
original gauge group $O(N_c+2)$ is broken to $O(N_c)$ with an additional $O(2)$ factor that decouples in the IR. Moreover, putting $Y'=-1$ into the superpotential we end up with $W=0$. Therefore, the residual theory is $O(N_c)$ with $N_f$ flavours.  The $R$-charge of the quarks in this theory is also equal to $R$; this is determined by the superpotential, which implies $N_f(1-R)-(N_c+2-2)=0$.   Thus, the
Hilbert series is $H^O_{N_c, N_f}(t, y, \vec x; R)$, in complete agreement with theory $A$.

There is also a similar duality for the special orthogonal gauge groups \cite{Aharony:2013kma}:
\paragraph{Theory $A'$:} $SO(N_c)$ with $N_f$ chirals in the vector representation and superpotential $W=Y$.
\paragraph{Theory $B'$:} $SO(N_f-N_c)$ with $N_f$ chirals in the vector representation, $N_f(2N_f+1)$ singlets $M$ and superpotential $W=M qq + \hat Y$. 
\\~\\
In theory $A'$, $Y=0$ in the chiral ring and so the Hilbert series of theory $A'$ is 
\be
	H^{(A')}(t, y, \vec x; R)=H^{SO}_{N_c, N_f}(t, y, \vec x; R)~,
\ee
where $H^{SO}_{N_c, N_f}(t, y, \vec x; R)$ is given by \eref{HSO}.  The moduli space of theory $B'$ can be conveniently studied by applying the ARSW duality and obtain 
$SO(N_c+2)$ with $N_f$ flavours, a singlet $\hat{Y}$ and $W=\hat{Y} \hat{Y}' +  \hat{Y}$, where $\hat{Y}'$ is the basic monopole operator in this theory.  The $F$-term $\partial_{\hat{Y}} W=0$ gives $\hat{Y}'=-1$, and so the gauge group $SO(N_c+2)$ is broken to $SO(N_c)$.  The residual theory is therefore $SO(N_c)$ gauge theory with $N_f$ flavours and zero superpotential.  The $R$-charge of the quarks in this theory is also equal to $R$.   Thus the Hilbert series of theory $B'$ is also $H^{SO}_{N_c, N_f}(t, y, \vec x; R)$.

\section{Dualities with quadratic monopole superpotentials} \label{sec:quadratic}
In this section, we study theories containing quadratic monopole superpotentials and their dualties.  We start our discussion with the duality involving models with superpotential $W = (X^+)^2 +  (X^-)^2 $, which was first proposed in \cite{Benini:2017dud}.  We then proceed to new dualities, including that involving $W=(X^-)^2$ and  chiral flavours and Chern--Simons levels, as well as those with symplectic and orthogonal gauge groups.  

Below we state explicitly the dualities, global symmetries, $R$-charges of various chiral fields, and some of their important features.  In section \ref{sec:S3part}, we present the three sphere partition functions of the theories study presented here and show that they match across the duality.  As a further test of these new dualties, it will also be shown that by giving appropriate real masses to certain chiral multiplets, we can flow from the proposed dualities to the known ones.  Subsequently, in section \ref{sec:HSquad}, we study the moduli spaces, compute the Hilbert series and show that they also match between the dual theories.

{
\subsection{The effect of quadratic monopole superpotential terms} \label{sec:effectquad}
The analysis is similar to the linear case.  If $V$ is one of the basic monopole operators, the superpotential $W=V^2$ fixes the $R$-charge of $V$ to be 1 and fixes the $R$-charge of the chiral fields.  { In the abelian theory, as pointed out in \cite{deBoer:1997kr, Dorey:1998kq}, we can rewrite $V$ as $V=e^z$, so that the $F$-term $\partial_z W= \partial_z e^{2z} =0$ implies that $e^{2z}=V^2=0$ and hence $V=0$; in other words, the Coulomb branch is lifted. We propose that this also holds in the non-abelian theory, namely the presence of the quadratic monopole operators in the superpotential also leads to the lift of the part of the Coulomb branch parametrised by that monopole operator.  We shall see from the analyses in section \ref{sec:HSquad} that this proposal is consistent with the duality.
}
}

\subsection{Models with unitary gauge groups}
\subsubsection{$W = (X^+)^2 +  (X^-)^2 $} \label{sec:XpsqXmsq}
\paragraph{Theory $A$:} $U(N_c)$ gauge theory with $N_f$ flavours $Q$ and $\tilde{Q}$, and superpotential
\be
W=(X^+)^2 + (X^-)^2~.
\ee
\paragraph{Theory $B$:} $U(N_f-N_c)$ gauge theory with $N_f$ flavours $q$ and $\tilde{q}$, $N_f^2$ singlets $M$ and superpotential 
\be W= M \tilde{q} q+ (\hat{X}^+)^2 + (\hat{X}^-)^2~.
\ee
\\
This duality was proposed in \cite{Benini:2017dud}.  Due to the monopole superpotential, we expect the $U(1)$ topological symmetry { to be broken to $\BZ_2$} and the $U(1)$ axial symmetry to be broken.  In theory $A$, the Coulomb branch is complete lifted due to the $F$-terms $\partial_{X^\pm} W=0$, which implies that $X^+ = X^- = 0$.  The same phenomenon happens also in theory $B$.  
The $R$-charge $R$ of the quarks and antiquarks in theory $A$ is fixed by the monopole superpotential:
\be \label{RXpsqXmsq}
N_f(1-R) -(N_c-1) = R[X^\pm]=1 \qquad \Rightarrow \qquad R= 1- \frac{N_c}{N_f}~.
\ee
The mesons in theory $A$ is mapped to the singlets $M$ in theory $B$.  We shall discuss the rank condition of $M$ and other details in \ref{sec:HSXpsqXmsq}.

\subsubsection{$W = (X^-)^2$} \label{sec:Xmsq}
\paragraph{Theory $A$:} $U(N_c)$ gauge theory with $N_f$ flavours $Q$ and $\tilde{Q}$ and $W=(X^-)^2$.
\paragraph{Theory $B$:} $U(N_f-N_c)$ gauge theory with $N_f$ flavours $q$ and $\tilde q$, $N_f^2$ singlets 
$\hat M$, singlet $S^+$ and superpotential $W=M \tilde{q} q + (\hat X^+)^2 + S^+ \hat X^- $ , where $\hat X^{\pm}$ are the basic monopoles in theory $B$, and $S^+$ is dual to the monopole $X^+$ in theory $A$. 
\\~\\
In these theories, the topological symmetry and the axial symmetry is broken to a diagonal subgroup, which we denotes by $U(1)_{T'}$ symmetry.  In theory $A$, the part of the Coulomb branch that is generated by $X^-$ is lifted { due to the quadratic term $(X^-)^2$ in the superpotential.}  However, the part that is generated by $X^+$ still remains.  The $R$-charges of $X^-$ is fixed to be equal to $1$, whereas that of $X^+$ depends on how we parametrise the mixing between the $R$-symmetry and the $U(1)_{T'}$ symmetry.  We shall postpone the detailed discussion until section \ref{sec:XmsqHS}.  We propose the following operator maps between theories $A$ and $B$:
\be
\begin{split}
\text{Theory $A$} & \qquad  \text{Theory $B$} \\
X^+ \quad &\longleftrightarrow \quad S^+~, \\
\text{mesons} \quad &\longleftrightarrow \quad M~.
\end{split}
\ee

In section \ref{sec:S3quadU}, we show that this duality can be obtained by flowing from the duality $W = (X^+)^2 +  (X^-)^2$ discussed in the previous subsection.  In this way, we also match the three sphere partition functions of the two theories in \eref{single}.  Furthermore, we demonstrate that, by appropriate shifts of real masses, one obtains the Aharony duality.  These constitute non-trivial tests of the proposed duality.

\subsubsection{$W = (X^-)^2$, chiral flavour and Chern--Simons terms} \label{sec:XmsqchflvCS}

\paragraph{Theory $A$:} $U(N_c)_{\frac{k}{2}}$ with $k>0$ and $(N_f, N_a=N_f-k)$ fund/antifund and superpotential $W=(X^-)^2$. 
\paragraph{Theory $B$:} $U(N_f-N_c)_{-k/2}$ with $(N_f, N_a=N_f-k)$ fund/antifund, $N_fN_a$ singlets and
superpotential $W=(\hat X^+)^2 + \sum_{i}^{N_f}\sum_{j}^{N_a}{M^i}_j \tilde{q}_i q^j$.
\\~\\
In theory $A$, The monopole operator $X^-$ vanishes in the chiral ring {due to the quadratic term $(X^-)^2$ in the superpotential,} and the Coulomb branch generated by $X^+$ is lifted due to the non-zero CS level.  Thus, the moduli space is generated by the meson matrix with the rank at most $N_c$. The $R$-charge $R$ of the fundamentals and antifundamentals is fixed by the monopole superpotential:
\be \label{RchargeXmsqCS}
1=  \frac{1}{2} N_f (1-R) +\frac{1}{2} N_a (1-R) -(N_c-1) \quad \Rightarrow \quad R =1-2 \frac{N_c}{N_a+N_f}~.
\ee
A similar analysis can be carried out for theory $B$, where singlets $M$ are mapped to the elements of the meson matrix of theory $A$.  We analyse the moduli space of theory $B$ in detail in section \ref{sec:XmsqCS}.   The contact terms and the matching of the three sphere partition functions are discussed in section \ref{sec:S3XmsqCS}.

\subsection{Models with symplectic gauge groups} \label{sec:sympY2}
\paragraph{Theory $A$:} $USp(2N_c)$ with $2N_f$ fundamentals and superpotential $W=Y^2$.
\paragraph{Theory $B$:} $USp(2(N_f-N_c-1))$ with $2N_f$ fundamentals, $N_f(2N_f-1)$ singlets $M$, and superpotential 
$W=M q q + \hat{Y}^2$. 
\\~\\
In theory $A$ the Coulomb branch is lifted.  Again, the moduli space is generated by the meson matrix with the rank at most $2N_c$.  The $R-$charge $R$ of the fundamentals is fixed by the monopole superpotential:
\be \label{RmonosympY2}
R[Y]=1= 2N_f(1-R) - 2N_c  \quad \Rightarrow \quad R = \frac{2N_f-2N_c-1}{2N_f}~.
\ee
A similar analysis can be carried out for theory $B$.  More details will be provided in section \ref{sec:HSsympY2}.  The matching of the partition functions for the two theories will be discussed in section \ref{sec:S3symp}.

\subsection{Models with orthogonal gauge groups} \label{sec:orthY2}
\paragraph{Theory $A$:} $O(N_c)$ with $N_f$ chirals in the vector representation and superpotential $W=Y^2$.
\paragraph{Theory $B$:} $O(N_f-N_c+2)$ with $N_f$ chirals in the vector representation, $N_f(2N_f+1)$ singlets $M$ and superpotential $W=M qq + \hat Y^2$. 
\\~\\
The analysis is very similar to the case of the symplectic gauge groups.  In theory $A$, the Coulomb branch is lifted.  Again, the moduli space is generated by the meson matrix with the rank at most $N_c$.  The $R-$charge $R$ of the quarks is fixed by the monopole superpotential:
\be \label{RmonoorthY2}
R[Y]=1= N_f(1-R)-(N_c-2)  \quad \Rightarrow \quad    R=\frac{N_f-N_c+1}{N_f}.
\ee
A similar analysis can be carried out for theory $B$.  More details will be provided in section \ref{sec:HSorthY2}.  The matching of the partition functions for the two theories will be discussed in section \ref{sec:S3orth}.

Finally, it is worth mentioning that there is also a similar duality for the special orthogonal gauge groups:
\paragraph{Theory $A'$:} $SO(N_c)$ with $N_f$ chirals in the vector representation and superpotential $W=Y^2$.
\paragraph{Theory $B'$:} $SO(N_f-N_c+2)$ with $N_f$ chirals in the vector representation, $N_f(2N_f+1)$ singlets $M$ and superpotential $W=M qq + (\hat Y)^2$.

\section{The three sphere partition function} \label{sec:S3part}
In this section we provide some analytic checks of the dualities 
proposed above.
We study the consistency of the real mass flows connecting the
dualities with quadratic monopole superpotentials in the UV
to Aharony duality in the IR.
In each case we conjecture in the UV an identity between the 
squashed three sphere partition functions obtained by localization.
Such identities are sketchily of the form 
\begin{equation}
\label{main}
 Z_{ele}(\mu) = Z_{mag}(\mu)
\end{equation}
where $Z_{ele}$ and $Z_{mag}$ refer to the electric and to the magnetic partition functions. The parameters $\mu$ are in general complex combinations of real masses and R-charges. The presence of 
the monopole superpotentials break some combinations of the topological and of the axial global symmetries and it reflects in some constraints
on the $\mu$ parameters. We refer to the these constraints as balancing conditions. 
Then we simulate, on the partition functions, the real mass flows that lead to other IR dualities. Such real mass flows correspond to infinite and real shifts on some of the parameters $\mu$.
In general we arrive, in the IR, to identities between infinite quantitites.
If we can drop the divergent terms in these final identities we interpret the
identities between the finite parts as the ones between the IR partition functions obtained after the real mass flows.
When the expected IR duality corresponds to Aharony duality we read the 
final identity between the electric and magnetic partition functions and compare with the ones that already appeared in the literature.
If these agree the whole procedure has furnished a consistency check of
(\ref{main}) and consequently of the conjectured UV duality.

We apply the procedure described above to the dualities with quadratic monopole superpotential in presence of unitary, symplectic and orthogonal gauge groups.
As a general remark we observe that in each case we need to perform a dual Higgs flow \cite{Intriligator:2013lca} in order to recover Aharony duality in the IR.
Such dual higgsing is necessary to reconstruct the correct scaling of the divergent term and to reconstruct the correct matter and gauge content of
the dual theories.

\subsection{The unitary case} \label{sec:S3quadU}

We start analyzing the RG flow from the $W = (X^+)^2 +  (X^-)^2$ duality
to Aharony duality.
As an intermediate step we obtain the duality with $W = (X^-)^2$.
Starting from this last duality we discuss the case with a CS term as well.

\subsubsection{Flowing from $W = (X^+)^2 +  (X^-)^2$ to $W = (X^-)^2$}

We start considering the duality between 3d $\mathcal{N}=2$ 
$U(N_c)$ with $N_f$ flavors and $W = (X^+)^2 +  (X^-)^2$ to $W = (X^-)^2$
and $U(\widetilde{N}_c = N_f-N_c)$ 
with $N_f$ dual flavors, the meson $M$ and 
$W = M q \tilde q + (\hat{X}^+)^2 + (\hat{X}^-)^2$.
{
The identity relating the UV partition functions
is of the form
}
\begin{equation}
\label{conj1}
Z_{U(N_c),N_f} (\mu, \nu)
=
\prod_{i,j=1}^{N_f} \Gamma_h(\mu_i+\nu_i)
Z_{U(\widetilde{N}_c),N_f} (\widetilde \mu, \widetilde \nu)
\end{equation}
where
\begin{equation}
Z_{U(N_c),N_f} (\mu,\nu) 
=
\frac{1}{|W|}
\int \prod_{a=1}^{N_c} \Big( d x_a
\prod_{i=1}^{N_f} \Gamma_h(\mu_i-x_a)
\Gamma_h(\nu_i+x_a) \Big)
\! \!\!\! 
\prod_{1\leq a < b \leq N_c}
\! \!\!\!
 \Gamma_h^{-1}(\pm(x_a-x_b))
\end{equation}
{
Even if this relation is so far conjectural, we will give 
an analytic proof below, observing that it can be derived 
from Aharony duality. 
}
The parameters $\mu_i$ and $\nu_i$ are complex combinations
of real masses and $R$-charges of the $N_f$ fundamentals and
anti-fundamentals respectively.  They can be explicitly expressed as follows
\begin{equation}
\mu_i = m_i + m_A + \omega \Delta,
\qquad
\nu_i = \widetilde m_i + m_A + \omega \Delta
\end{equation}
where $\sum_i  m_i = \sum_i  \widetilde{m}_i =0$.
The real parameter $m_A$ is the axial mass and 
the $R$-charge $\Delta$ coincides in this case for the fundamentals and
for the anti-fundamentals. The parameter $\omega$ is purely imaginary, and it is explicitly given by
$2 \omega \equiv \omega_1+\omega_2  \equiv  i (b + b^{-1})$, where $b$ is the real squashing parameter
of the ellipsoid $S_b^3$.  The ellipsoid is defined by the relation
\begin{equation}
\frac{x_1^2+x_2^2}{b^2}+\frac{x_3^2+x_4^2}{1/b^2}=1
\end{equation}
The one loop determinants $\Gamma_h$ are hyperbolic Gamma functions
\footnote{The notations adopted by this paper is related to that adopted by \cite{Benini:2017dud} as follows: 
\bea
\omega &= i\frac{Q}{2}~,  \quad \text{with}~ Q =b+b^{-1}~, \nn \\
\Gamma_h(x) &= s_b\left(i \frac{Q}{2} - x \right)~, \quad s_b(x)= \Gamma_h \left(\omega -x \right)~. \nn
\eea
Moreover, the notations of real masses in this paper are related to \cite{Benini:2017dud} as follows:
\be \nn
\begin{split}
\text{Ours} \quad&\qquad \quad \text{Ref. \cite{Benini:2017dud}} \\
\mu_a  \quad &\longleftrightarrow \quad  m_a \\
\nu_a   \quad &\longleftrightarrow \quad \tilde{m}_a \\
\frac{1}{2}(\mu_a+\nu_a)  \quad &\longleftrightarrow \quad  \mu_a \\
\frac{1}{2}(-\mu_a+\nu_a)  \quad &\longleftrightarrow \quad M_a~. \\
\end{split}
\ee
}
\begin{equation}
  \label{eq:Gammahvbd}
  \Gamma_h(z;\omega_1,\omega_2) \equiv
  \Gamma_h(z)\equiv 
  e^{
    \frac{i \pi}{2 \omega_1 \omega_2}
    ((z-\omega)^2 - \frac{\omega_1^2+\omega_2^2}{12})}
  \prod_{\alpha=0}^{\infty} 
  \frac
  {1-e^{\frac{2 \pi i}{\omega_1}(\omega_2-z)} e^{\frac{2 \pi i \omega_2 \alpha}{\omega_1}}}
  {1-e^{-\frac{2 \pi i}{\omega_2} z} e^{-\frac{2 \pi i \omega_1 \alpha}{\omega_2}}}.
\end{equation}
The dimension of the Weyl group is denoted by $|W|$.
In (\ref{conj1}) the real mass parameters are constrained by the 
balancing condition
\begin{equation}
\sum_{i=1}^{N_f} (\mu_i + \nu_i) = 2 \omega (N_f-N_c)
\end{equation}
that corresponds to the constraint enforced by the 
monopole superpotential on the global symmetries\footnote{Let us explain briefly the origin of this balancing condition. The superpotential $W=(X^+)^2+(X^-)^2$ constraints the $R$-charges of $X^\pm$ to be $1$, \ie~ $R[X^\pm]=1=N_f(1-\Delta)-(N_c-1)$, which implies that $\Delta= 1- \frac{N_c}{N_f}$. This is in agreement with the matching of the coefficients of $\omega$ in the left and the right hand sides of the balancing condition, namely $2N_f \Delta = 2(N_f-N_c)$.}.

{
As anticipated above the identity
\eref{conj1}
 can be derived from 
the one relating two Aharony dual theories.
The argument works as follows.
The identity for Aharony duality is  
\begin{eqnarray}
 \label{eq:Aharony}
Z_{U(N_c),N_f}(\mu;\nu;\eta)
&=&
\Gamma_h \left(\pm \frac{\eta}{2} - \frac{1}{2} \sum_{i=1}^{N_f} (\mu_i +\nu_i) + \omega (N_f-N_c+1) \right)
\\
&\times&
\prod_{i,j=1}^{N_f}
\Gamma_h(\mu_i + \nu_j)
Z_{U(N_f-N_c),N_f}(\omega-\mu;\omega-\nu;-\eta)
\nonumber
\end{eqnarray}
where the parameters $\mu$, $\nu$ and the FI term $\eta$ are unconstrained.
From this identity one can prove \eref{conj1}, 
with the help of a field theoretical analysis:
deforming the electric side of Aharony duality by the quadratic 
superpotential $ W = X_+^2 + X_-^2$
imposes the constraints
 \begin{equation}
 \label{eq:newconstraints}
\eta = 0,
\quad
 \sum_{i=1}^{N_f} (\mu_i +\nu_i)  = 2 \omega(N_f-N_c)
\end{equation}
By plugging (\ref{eq:newconstraints}) in the identity (\ref{eq:Aharony})
and by using the fact that $\Gamma_h(\omega) = 1$ 
one arrives at the identity \eref{conj1}.
A similar argument can be repeated for the cases of dualities 
with symplectic and orthogonal gauge groups. We leave the details to the reader.
}

We now consider this duality for $N_f+1$ fundamentals and anti-fundamentals. The  gauge group of the dual model is $U(N_f-N_c+1)$.
We study the real mass flow to the $W = (X^-)^2$ duality by shifting 
$\mu_{N_f+1}$ and $\nu_{N_f+1}$ as
\begin{equation}
\mu_{N_f+1} \rightarrow \frac{\eta}{2} + s,
\quad
\nu_{N_f+1} \rightarrow \frac{\eta}{2} - s
\end{equation}
and consider the limit $s \rightarrow \infty$.
The flow on the dual side must be supported by the Higgs flow
$x_{N_f-N_c+1} \rightarrow y + s$.
The balancing condition becomes
\begin{equation} \label{balancingX2}
\eta + \sum_{i=1}^{N_f}(\mu_i + \nu_i) = 2 \omega(N_f-N_c+1)
\end{equation}

At large $s$ we can integrate out the massive flavors by using the formula
\begin{equation}
\label{intout}
\lim_{x\rightarrow \infty} \Gamma_h(x) = e^{\frac{i \pi}{2}(x-\omega)^2}
\end{equation}
This formula corresponds to the generation of half-integer CS levels, 
for the gauge and for the flavors symmetries,
when integrating out heavy fermions with large real mass.
By using (\ref{intout}) the partition function of the electric theory becomes
\begin{equation}
e^{i \pi  N_c (\eta -2 \omega )t}
\int \prod_{a=1}^{N_c} e^{- i \pi(\eta-2 \omega)x_a}
 dx_a \prod_{i=1}^{N_f} \Gamma_h(\mu_i-x_a)
\Gamma_h(\nu_i+x_a) 
\! \!\!\! \! 
\prod_{1\leq a<b \leq N_c}
\! \!\!\!\! 
 \Gamma_h^{-1}(\pm(x_a-x_b))
\end{equation}
where the exponential factor in the integrand represents the
contribution of a generalized FI term in the classical action.
The dual partition function is
\begin{eqnarray}
\label{dual}
&&e^{i \pi  N_c (\eta -2 \omega )t}
\Gamma_h(\eta)
\prod_{i,j=1}^{N_f} \Gamma_h (\mu_i+\nu_j)
\int d y e^{-i \pi y (\eta -2 \omega )}
\Gamma_{h} \Big(\pm x+\omega -\frac{\eta}{2}\Big)
\\
&&
\int \prod_{a=1}^{N_f-N_c} e^{- i \pi(\eta-2 \omega)x_a}
d x_a
\prod_{i=1}^{N_f} \Gamma_h(\omega-\mu_i-x_a)
\Gamma_h(\omega-\nu_i+x_a) 
\! \!\!\! \!\! \!\!\!\! 
\prod_{1\leq a<b \leq \tilde N_f-N_c}
\! \!\!\!\! \! \!\!\!\! 
 \Gamma_h^{-1}(\pm(x_a-x_b))
 \nonumber 
 \end{eqnarray}
The first integral in (\ref{dual}) can be simplified, 
because it corresponds to SQED with 
one flavor, and it is mirror dual to a triple of singlets.
This duality corresponds to the integral identity
\begin{equation}
\int d x e^{i \pi  x \lambda}
\Gamma_{h} (x+m) \Gamma_{h} (x+n)
=
\Gamma_h\left( m+n\right)
\Gamma_h\left(\pm \frac{\lambda}{2}-\frac{m+n}{2}+\omega\right)
\end{equation}
In our case it corresponds to 

\begin{equation}
\int d x e^{-i \pi  x (\eta -2 \omega )}
\Gamma_{h} \Big(\pm x+\omega -\frac{\eta}{2}\Big)
=
\Gamma_h\left( 2\omega-\mu\right)
\Gamma_h\left(\pm \frac{2\omega-\eta}{2}+\frac{\eta}{2}\right)
\end{equation}
Then, substituting this integral in the magnetic partition function and using  the identities
\begin{equation}
\Gamma_h(2\omega-x) \Gamma_h(x) =1~,\qquad
\Gamma_h(\omega)=1
\end{equation}
 we arrive at the relation
\begin{eqnarray}
\label{single}
&&
\int \prod_{a=1}^{N_c} dx_a
e^{i \pi(2 \omega-\eta) x_a}
\prod_{i=1}^{N_f} \Gamma_h(\mu_i-x_a)
\Gamma_h(\nu_i+x_a) 
\! \!\!\! \! \!\!\!
\prod_{1\leq a<b \leq N_c}
\! \!\!\!\! \!\!\!
 \Gamma_h^{-1}(\pm(x_a-x_b)) 
   \nonumber 
\\
=
 &&
 \Gamma_h(\eta-\omega)
\prod_{i,j=1}^{N_f} \Gamma_h (\mu_i+\nu_j)
\int \prod_{a=1}^{N_f-N_c}  d x_a
e^{ i \pi(2 \omega-\eta)x_a}
   \nonumber 
\\
\times 
 &&
\prod_{i=1}^{N_f} \Gamma_h(\omega-\mu_i-x_a)
\Gamma_h(\omega-\nu_i+x_a) 
\! \!\!\! \! \!\!\!
\prod_{1\leq a<b \leq N_f-N_c}
\! \!\!\!\! \!\!\!
 \Gamma_h^{-1}(\pm(x_a-x_b))
\end{eqnarray}
The first term on the RHS corresponds to the singlet $S$.  It can be re-written, by using the balancing condition,
as 
\begin{equation}
\begin{split}
\label{fourteen}
 \Gamma_h(\eta-\omega)
 &=
  \Gamma_h \left( \frac{\eta}{2}
  -\frac{1}{2} \sum_{i=1}^{N_f} (\mu_i+\nu_i) +\omega (N_f-N_c)
   \right)~.
\end{split}   
\end{equation}

{
At this point of the discussion we can shift $\eta \rightarrow \eta+2 \omega$, 
so that now $-\frac{1}{2}\eta$ can be interpreted as a real FI parameter of the 
theory, as it appears in the exponential functions in the first and the second lines of \eref{single}. 
This shift modifies the balancing condition \eref{balancingX2} to
\be
\label{bc2}
\eta+\sum_{i=1}^{N_f} (\mu_i +\nu_i) = 2 \omega(N_f-N_c)
\ee
and the contribution of the singlet in \eref{fourteen} becomes
\be
\label{fourteen2}
  \Gamma_h \left( \frac{\eta}{2}
  -\frac{1}{2} \sum_{i=1}^{N_f} (\mu_i+\nu_i) +\omega (N_f-N_c+1)
   \right)~.
\ee
We can then read the charge of the singlet $S$ from the argument of the 
hyperbolic Gamma function appearing in \eref{fourteen2} 
and then relate it to the one of the electric monopole $X^+$.  
The term $\frac{\eta}{2}$ implies that $S$ carries the topological charge $+1$, in the same way as $X^+$ does. 
The axial mass can be read from the sum of the masses $\mu_i$ and $\nu_i$.
This is because each of these masses can be split into a vector and an axial contribution, and the sum corresponds 
to the axial contribution only. 
More explicitly we can define $\mu_i$ and $\nu_i$ as
\be
\label{massesmunu}
\mu_i = m_i + m_A + \omega \Delta,
\quad
\nu_i = n_i + m_A + \omega \Delta,
\quad \text{with} \quad
\sum_{i=1}^{N_f} m_i = \sum_{i=1}^{N_f} n_i = 0 
\ee
where $m_A$ is the axial mass and $\Delta$ refers to the $R$-charge.
This axial mass of the singlet is $-m_A$ and it corresponds to the one of $X^+$.
Observe that the presence of quadratic monopole superpotentials actually preserves only 
a linear combination of the axial and of the topological 
symmetries. The role of such a superpotential deformation is here played by the 
balancing condition \eref{bc2}.

Eventually we can match the $R$-charges of the singlet $S$ and of the monopole $X^+$.
The $R$-charge of $S$ corresponds to the coefficient of $\omega$ in the argument of 
\eref{bc2} after the substitution \eref{massesmunu}.
It is 
\be
R[S] = N_f(1-\Delta) - N_c+1
\ee
Thus, $R[X^+]=R[S]$, as expected.

The final relation \eref{single} is therefore compatible with the duality
between $U(N_c)$ with $N_f$ fundamental flavors  and $W = (X^-)^2$ and
$U(N_f-N_c)$ with $N_f$ dual fundamental flavors and $W = M q 
\tilde q +(\hat{X}^+)^2+S \hat{X}^-$, where 
the singlet $S$ corresponds to $X^+$ in the electric side.  The term $S\hat{X}^-$, which enters the superpotential, has $R$-charge $2$.
This follows from the fact that 
the $R$-charge of $\hat{X}$ is given by
\be
R[\hat{X}] = N_f(1-(1-\Delta)) - (N_f - N_c-1)~,
\ee
and so $R[S] + R[\hat{X}] =2$.
}

\subsubsection{Flowing from $W = (X^-)^2$ to the Aharony duality}
We can study a further flow, from this duality to the Aharony duality.
This flow is engineered by considering the $W = (X^-)^2$ duality with 
with $N_f+1$ flavors and
shift the masses as
\begin{eqnarray}
(\mu_i,\nu_i) &\rightarrow& (\mu_i-s,\nu_i+s)
\quad \quad \quad \quad
i=1,\dots,N_f
\\
(\mu_{N_f+1},\nu_{N_f+1}) 
&\rightarrow& 
\big(\frac{\eta_2}{2}+s N_f,\frac{\eta_2}{2}-s N_f \big)
\end{eqnarray} 

We also shift the vector multiplet by $x_i \rightarrow x_i+s$,
$i=,\dots,N_c$.
Furthermore, in the dual theory we need to consider 
the shift
 $x_i \rightarrow x_i-s$, for $i=1,\dots,N_f-N_c$. There is also a dual Higgsing corresponding to the shift
$x_{N_f-N_c+1}\rightarrow y+ s N_f$.
The balancing condition \eref{balancingX2} becomes 
\begin{equation}
\label{bb}
\eta+\eta_2 + \sum_{i=1}^{N_f}(\mu_i + \nu_i) = 2 \omega(N_f-N_c+2)~.
\end{equation}
By computing the large $s$ limit we arrive at the identity

\begin{eqnarray}
\label{rel}
&&
\int \prod_{a=1}^{N_c} 
dx_a
e^{- i \pi(\eta-\eta_2)x_a}
\prod_{i=1}^{N_f} \Gamma_h(\mu_i-x_a)
\Gamma_h(\nu_i+x_a) 
\! \!\!\! \! 
\prod_{1\leq a<b \leq N_c}
\! \!\!\!\! 
 \Gamma_h^{-1}(\pm(x_a-x_b)) 
  \nonumber \
  \\
= &&
  \Gamma_h(\eta-\omega)
 \Gamma_h(\eta_2)
\prod_{i,j=1}^{N_f} \Gamma_h (\mu_i+\nu_j)
\int 
\prod_{a=1}^{N_f-N_c} 
dx_a
e^{- i \pi(\eta-\eta_2)x_a}
  \nonumber \\
 \times &&
\prod_{i=1}^{N_f} \Gamma_h(\omega-\mu_i-x_a)
\Gamma_h(\omega-\nu_i+x_a) 
\! \!\!\! \! \
\prod_{1\leq a<b \leq N_f- N_c}
\! \!\!\!\! 
 \Gamma_h^{-1}(\pm(x_a-x_b))
  \nonumber \\
 \times   &&
\int dy
e^{-i \pi  y (\sum _{i=1}^{N_f} (\mu _i+ \nu _i)
+\eta -2 \omega(N_f-N_c+1)}
\Gamma _h \big(y+\omega-\frac{\eta_2}{2}\big)
\Gamma _h \big(-y+\omega-\frac{\eta_2}{2}\big)
\end{eqnarray}
The integral in the last line can be computed 
explicitly and it corresponds to 
\begin{eqnarray}
\label{abe}
&&
\Gamma_h(2 \omega-\eta_2)
\Gamma _h\Big(\frac{\eta_2}{2} + \big(
\frac{\eta}{2}+ \omega(N_f-N_c+1)+\frac{1}{2}\sum _{a=1}^{N_f} (\mu _a+ \nu _a)
\big)\Big)
\nonumber \\
\times &&
\Gamma _h\Big(\frac{\eta_2}{2} - \big(
\frac{\eta}{2} { -} \omega(N_f-N_c+1)+\frac{1}{2}\sum _{a=1}^{N_f} (\mu _a+ \nu _a)
\big)\Big)
\end{eqnarray}
The first term in (\ref{abe}) simplifies with the term $\Gamma_h (\eta_2)$ in the first line of (\ref{rel}), due to the identity $\Gamma_h(2 \omega- \eta_2) \Gamma_h (\eta_2) = 1$.
The second term is equivalent to $\Gamma_h(\omega) = 1$ because of the balancing condition
(\ref{bb}).
The last term in (\ref{abe}), which is simplified to $\Gamma_h(\eta_2-\omega)$ upon using the balancing condition \eref{bb}, can be identified with the monopoles of the electric theory. The antimonopole
is still identified with $\Gamma_h(\eta-\omega)$ in  (\ref{rel}).
Indeed by using the balancing condition \eref{bb} we can see that they are  equivalent to 
\begin{equation}
\begin{split}
 \Gamma_h(\eta_2-\omega) &= \Gamma _h\Big( \frac{\eta_2-\eta}{2} { +} \omega(N_f-N_c+1)
-\frac{1}{2}\sum _{a=1}^{N_f} (\mu _a+ \nu _a) \Big)\\ 
 \Gamma_h(\eta-\omega)   &= \Gamma _h\Big(- \frac{\eta_2-\eta}{2} { +} \omega(N_f-N_c+1)
-\frac{1}{2}\sum _{a=1}^{N_f} (\mu _a+ \nu _a)
\Big)~,
\end{split}
\end{equation}
where $\eta_2-\eta \equiv \zeta$ is the effective FI that can be read from the partition function.
We have obtained the identity
\begin{eqnarray}
&&
\int \prod_{a=1}^{N_c} 
dx_a
e^{- i \pi \zeta x_a}
\prod_{i=1}^{N_f} \Gamma_h(\mu_i-x_a)
\Gamma_h(\nu_i+x_a) 
\! \!\!\! \! 
\prod_{1\leq a<b \leq N_c}
\! \!\!\!\! 
 \Gamma_h^{-1}(\pm(x_a-x_b)) =
 \nonumber \\
 &&
 \Gamma _h\Big(\pm \frac{\zeta}{2} { +} \omega(N_f-N_c+1)
-\frac{1}{2}\sum _{a=1}^{N_f} (\mu _a+ \nu _a)
\Big)
 \prod_{i,j=1}^{N_f} \Gamma_h (\mu_i+\nu_j)
 \\
 \times 
  && 
\int 
\prod_{a=1}^{N_f-N_c} 
dx_a
e^{- i \pi \zeta x_a}
\prod_{i=1}^{N_f} \Gamma_h(\omega-\mu_i-x_a)
\Gamma_h(\omega-\nu_i+x_a) 
\! \!\!\! \! \! \!\!\! 
\prod_{1\leq a<b \leq N_f- N_c}
\! \!\!\! \! \! \!\!\! 
 \Gamma_h^{-1}(\pm(x_a-x_b))
  \nonumber 
 \end{eqnarray}
 This is the correct expression for the matching 
 of the electric and the magnetic partition
 function in the Aharony duality.
 
 Summarizing we started from the conjectured identity
 (\ref{conj1}) between the partition functions of the $W = (X^+)^2+(X^-)^2$
 duality.
 Then we have obtained the identity between the 
 partition functions of the $W =(X^-)^2$ duality.
 Eventually we have obtained the known identity  
corresponding to the matching between the electric and the magnetic
Aharony dual phases.
This corroborates the validity of the dualities with quadratic monopole superpotentials.

\subsubsection{The $W= (X^-)^2$ duality with the Chern-Simons term} \label{sec:S3XmsqCS}
We conclude this section by studying the RG flow 
from the $W= (X^-)^2$ duality to the case with CS term.
The masses, the FI and the scalar $\sigma$ in the electric theory are shifted as follows
\begin{eqnarray}
\begin{array}{ll}
\mu_a \rightarrow \mu_a  - k s  & \quad a=1,\dots,N_f-k \\
\mu_a \rightarrow \mu_a + (2 N_f-k) s 
&\quad a=N_f-k+1,\dots,N_f \\
\nu_a \rightarrow \nu_a  + k s  &\quad a=1,\dots,N_f \\
\eta \rightarrow \eta - 2 N_f k s \\
\sigma_i \rightarrow \sigma_i-k s& \quad  i=1,\dots,N_c \\
\end{array}
\end{eqnarray}
While in the magnetic theory we read the masses from the duality map and
provide the opposite shift on $\sigma_i \rightarrow \sigma_i + k s$,
$i=1,\dots,\widetilde{N}_c$.
 We arrive to a duality between a $U(N_c)_{\frac{k}{2}}$ theory with $N_f-k$ fundamentals
and $N_f$ antifundamentals
and superpotential $W = (X^+)^2$
and a $U(\widetilde{N}_c=N_f-N_c)_{-\frac{k}{2}}$
theory with $N_f-k$ fundamentals
and $N_f$ antifundamentals, a meson $M$ with 
$N_f(N_f-k)$ components and superpotential
\begin{equation}
W = \sum_{i=1}^{N_f} \sum_{j=1}^{N_f-k} M_i^j q_j \tilde q_i + (\hat{X}^-)^2
\end{equation}
We perform the infinite shift on the identity (\ref{single}).
There is a divergent phase that cancels between the electric
and the magnetic side of the identity.
Observe that the linear divergent term in the phase cancels because of the relation
\begin{equation}
\label{usefulerel}
\sum_{i=1}^{N_f} \mu_i =\sum_{i=1}^{N_f-k} \mu_i + 
\sum_{i=N_f-k+1}^{N_f} \mu_i 
=
\sum_{i=1}^{N_f} \nu_i 
\end{equation}
The final identity is 
\begin{eqnarray}
&&
\frac{1}{|W_{U(N_c)}|}
\int \prod_{a=1}^{N_c} d \sigma_a e^{i  \pi \xi_e \sigma_a + \frac{i \pi}{2} k \sigma_a^2}
\prod_{i=1}^{N_f-k} \Gamma_h(\mu_i + \sigma_a) 
\prod_{j=1}^{N_f} \Gamma_h(\nu_j - \sigma_a) 
\nonumber \\
\times
&&
\prod_{a<b} \Gamma_h^{-1}(\pm(\sigma_a - \sigma_b) )
=
e^{-i \pi \phi}
\prod_{i=1}^{N_f-k} \prod_{j=1}^{N_f}  \Gamma_h(\mu_i + \nu_j)
\frac{1}{|W_{U(\widetilde N_c)}|}
\int 
\prod_{a=1}^{\widetilde{N}_c} d \sigma_a e^{i  \pi \xi_m \sigma_a - \frac{i \pi}{2} k \sigma_a^2}
\nonumber \\
\times
&&
\prod_{i=1}^{N_f-k} \Gamma_h(\omega-\mu_i + \sigma_a) 
\prod_{j=1}^{N_f} \Gamma_h(\omega-\nu_j - \sigma_a) 
\prod_{a<b} \Gamma_h^{-1}(\pm(\sigma_a - \sigma_b))
\nonumber 
\end{eqnarray}
where the phase corresponding to the contributions of the
contact terms \cite{Benini:2011mf, Closset:2012vg, Closset:2012vp} is
\begin{eqnarray}
\phi  
&=&
\Big(\sum _{i=1}^{N_f} \nu _i + \omega  (N_c-N_f)\Big)
 \Big( \sum _{i=1}^{N_f-k} \mu _i-\sum _{i=1}^{N_f} \nu _i \Big)
+k \omega  \sum _{i=1}^{N_f} \nu _i
\nonumber \\
&&
-\frac{1}{2} k \sum _{i=1}^{N_f} \nu _i^2
+\frac{1}{2} k \omega ^2 (N_c-N_f)
+\frac{1}{2} (\eta-\omega)^2
\end{eqnarray}
and the effective electric and magnetic FI terms are
\begin{eqnarray}
\xi_e &=& -\sum _{i=1}^k \mu _i-\eta +k \omega
=
\sum _{i=1}^{N_f-k} \mu _i-\sum _{i=1}^{N_f} \nu _i-\eta +k \omega +\omega 
\nonumber \\
\xi_m &=& -\sum _{i=1}^k \mu _i-\eta 
=
\sum _{i=1}^{N_f-k} \mu _i-\sum _{i=1}^{N_f} \nu _i-\eta +\omega
\end{eqnarray}
where in the last equalities we made use of the balancing condition (\ref{balancingX2}) and  of the relation (\ref{usefulerel}).

\subsection{The symplectic case} \label{sec:S3symp}

In this section we perform a similar analysis for the 
duality between a 3d $\mathcal{N}=2$  $USp(2N_c)$ 
theory with $2 N_f$ fundamentals and superpotential
$W = Y^2$ and an $USp(2(N_f-N_c-1))$ theory with 
$2 N_f$ fundamentals, an anti-symmetric meson 
and superpotential $W = M q q + \hat Y^2$. 
Observe that a similar duality with a linear monopole superpotential
has already been studied in the literature 
\cite{Aharony:2013dha} and it corresponds
to the duality obtained by circle reduction of 4d $USp(2 N_c)$ Seiberg duality
\cite{Intriligator:1995ne}.

Assuming the validity of this duality 
the identity between the partition functions 
on $S_b^3$ is
\begin{eqnarray}
\label{SPquad}
\frac{1}{2^{N_c} N_c!}
\int \prod_{a=1}^{N_c} d x_a
\frac{\prod_{i=1}^{2 N_f} \Gamma_h(\pm x_a + \mu_i)}
{\Gamma_h(\pm 2 x_a)}
\!\!\!\!\!
\prod_{1\leq a<b \leq N_c} 
\!\!\!
\frac{1}{\Gamma_h(\pm x_a \pm x_b)}
=\!\!\!\prod_{1 \leq i<j \leq N_f} 
\!\!\!\Gamma_h(\mu_i + \mu_j)
&&
\nonumber 
\\
\frac{1}{2^{\widetilde N_c} \widetilde N_c!}
\int \prod_{a=1}^{\widetilde N_c} d x_a
\frac{\prod_{i=1}^{2 N_f} \Gamma_h(\pm x_a +\omega- \mu_i)}{\Gamma_h(\pm 2 x_a)}
\prod_{1\leq a<b \leq \widetilde N_c} 
\frac{1}{\Gamma_h (\pm x_a \pm x_b)}
&&
\nonumber 
\\
\end{eqnarray}
with the balancing condition
\begin{equation}
\sum_{i=1}^{2N_f} \mu_i =  \omega (2 N_f-2N_c-1)
\end{equation}
As a check we show that (\ref{SPquad})
becomes the identity between the partition functions of the
electric and of the magnetic phases of Aharony duality.
We trigger the real mass flow by 
performing the shifts
\begin{equation}
\mu_{2 N_f+1} = \frac{\eta}{2} + s
\quad
\mu_{2 N_f+2} =\frac{\eta}{2}  - s
\end{equation}
where we impose $s$ to be large and positive.
In the dual theory we need to perform an higgsing as well,
$\widetilde \sigma_{N_f-N_c} \rightarrow y+s$.
The balancing condition becomes
\begin{equation}
2 \eta + \sum_{i=1}^{2N_f} \mu_i =  \omega (2 N_f-2N_c+1)
\end{equation}
One can show that the divergent term coincide and arrive to the identity
\begin{eqnarray}
\label{ddd}
&&
\frac{1}{2^{N_c} N_c!}
\int \prod_{a=1}^{N_c} d x_a
\frac{\prod_{i=1}^{2 N_f} \Gamma_h(\pm x_a + \mu_i)}{\Gamma_h(\pm(2 x_a))}
\prod_{1\leq a<b \leq N_c} \frac{1}{\Gamma_h(\pm x_a \pm x_b)}
\nonumber 
\\
=
&&
\Gamma_h(2 \alpha) 
\prod_{1 \leq i<j \leq 2 N_f} \Gamma_h(\mu_i + \mu_j)
\frac{1}{2^{\widetilde N_c} \widetilde N_c!}
\int \prod_{a=1}^{\widetilde N_c} d  x_a
\frac{\prod_{i=1}^{2N_f} \Gamma_h(\pm x_a +\omega- \mu_i)}{\Gamma_h(\pm 2 x_a)}
\nonumber 
\\
\times
&&
\prod_{1\leq a<b \leq \widetilde N_c} 
\frac{1}{\Gamma_h(\pm x_a \pm x_b)}
\int
dy e^{2 \pi i (\alpha-\omega) y}
\Gamma_h(\pm y + \alpha)
\end{eqnarray}
where $\widetilde N_c = N_f-N_c-1$.
This is a duality between a $USp(2N_c)$ theory
and a $USp(2 (N_f-N_c-1)) \times U(1)$
theory.
In order to arrive at a more conventional duality we can reformulate this
identity by integrating over the $U(1)$ factor.
The last integral in (\ref{ddd}) is equivalent to the product
\begin{equation}
\label{produ}
\Gamma_h(2 \alpha-\omega) 
\Gamma_h(\omega)
\Gamma_h(2 \omega-2 \alpha)
\end{equation}
The second term in (\ref{produ})
 is exactly equal to $1$ while the 
last term  in (\ref{produ}) simplifies against $\Gamma_h(2 \alpha)$
in (\ref{ddd}).
We are left with the term $\Gamma_h(2 \alpha-\omega) $.
By applying the balancing condition 
this is equivalent to 
\begin{equation}
\Gamma_h\Big(2 \omega(N_f-N_c) - \sum_{i=1}^{2N_f} \mu_i \Big)
\end{equation}
that is the contribution of the electric  monopole
acting as a singlet in the dual phase.
The final identity is
\begin{eqnarray}
\label{ddd}
&&
\frac{1}{2^{N_c} N_c!}
\int \prod_{a=1}^{N_c} d x_a
\frac{\prod_{i=1}^{2 N_f} \Gamma_h(\pm x_a + \mu_i)}{\Gamma_h(\pm(2 x_a))}
\prod_{1\leq a<b \leq N_c} \frac{1}{\Gamma_h(\pm x_a \pm x_b)}
\nonumber 
\\
=
&&
\Gamma_h\Big(2 \omega(N_f-N_c) - \sum_{i=1}^{2N_f} \mu_i \Big)
\prod_{1 \leq i<j \leq 2 N_f} \Gamma_h(\mu_i + \mu_j)
\frac{1}{2^{\widetilde N_c} \widetilde N_c!}
\nonumber 
\\
\times
&&
\int \prod_{a=1}^{\widetilde N_c} d  x_a
\frac{\prod_{i=1}^{2N_f} \Gamma_h(\pm x_a +\omega- \mu_i)}{\Gamma_h(\pm 2 x_a)}
\prod_{1\leq a<b \leq \widetilde N_c} 
\frac{1}{\Gamma_h(\pm x_a \pm x_b)}
\end{eqnarray}
and it represents again the matching of the partition functions 
of Aharony duality for $USp(2N_c)$ gauge theories.

\subsection{The orthogonal case} \label{sec:S3orth}

In this section we conclude the analysis by studying the 
duality between a 3d $\mathcal{N}=2$ a $O(N_c)$ 
theory with $N_f$ vectors 
and superpotential
$W = Y^2$ and a $O(N_f-N_c+2)$ theory with 
$N_f$ dual vectors, an symmetric meson 
and superpotential $W = M q q + \hat Y^2$. 
Observe that a similar duality with a linear monopole superpotential
has already been studied in the literature 
\cite{Aharony:2013kma} and it corresponds
to the duality obtained by circle reduction of 4d $O(N_c)$ Seiberg duality
\cite{Intriligator:1995id}. 

We will not specify the global properties in this discussion, meaning that we analyze the duality between $O(N_c)_+$ theories in the language of \cite{Aharony:2013kma}.  It should be nevertheless interesting to study the other cases, because in this case the different global properties are expected to have observable effects on the models and consequently on the dualities. We leave this problem for future analysis.

%

At the level of the partition function wee need to distinguish the 
even and odd $N_c$ case.
In the first case, if we consider an $O(2N_c)$ theory with $2 N_f$ fundamentals the duality corresponds to the identity
\begin{eqnarray}
&&
\frac{1}{2^{N_c-1}N_c!}
\int \prod_{a=1}^{N_c} d x_a
\prod_{i=1}^{2 N_f} \Gamma_h(\pm x_a + \mu_i)
\prod_{1\leq a<b \leq N_c} \frac{1}{\Gamma_h(\pm x_a \pm x_b)}
=\prod_{1 \leq i \leq j \leq 2 N_f} \Gamma_h(\mu_i+ \mu_j)
\nonumber \\
&&
\frac{1}{2^{\widetilde{N_c}-1} \widetilde{N_c}!}
\int \prod_{i=1}^{\widetilde{N_c}} d x_a
\prod_{a=1}^{2 N_f} \Gamma_h(\pm x_a +\omega- \mu_i)
\prod_{1\leq a<b \leq \widetilde N_c} \frac{1}{\Gamma_h(\pm x_a \pm x_b)}
\end{eqnarray}
with the balancing condition
\begin{equation}
\sum_{i=i}^{2N_f} \mu_i=  \omega (2N_f-2N_c+1)
\end{equation}
If the rank is odd, corresponding to a $O(2N_c+1)$ theory,
the identity is
\begin{eqnarray}
&&
\frac{\prod_{i=1}^{2N_f} \Gamma_h(\mu_i)}{2^{N_c}N_c!}
\int \prod_{a=1}^{N_c} d x_a
\frac{
\prod_{i=1}^{2N_f} \Gamma_h(\pm x_a+ \mu_i)
}
{\Gamma_h(\pm x_a)}
\!\!\!\!
\prod_{1\leq a<b \leq N_c} \frac{1}{\Gamma_h(\pm x_a \pm x_b)}
=
\!\!\!\!\!\!\!\!\!
\prod_{1 \leq i \leq j \leq 2 N_f} 
\!\!\!\! \!\!\!\!
\Gamma_h(\mu_i + \mu_j)
\nonumber \\
\times 
&&
\frac{
\prod_{i=1}^{2 N_f} \Gamma_h(\omega-\mu_i)
}{2^{\widetilde{N_c}} \widetilde{N_c}!}
\int \prod_{a=1}^{\widetilde{N}} d x_a
\frac{
\prod_{i=1}^{2N_f} \Gamma_h(\pm x_a +\omega- \mu_i)
}
{\Gamma_h(\pm x_a)}
\!\!\!\!
\prod_{1\leq a<b \leq \widetilde N_c} 
\!\!\!\!
\frac{1}{\Gamma_h(\pm x_a 
\pm x_b)}
\end{eqnarray}
with the balancing condition
\begin{equation}
\sum_{i=1}^{2 N_f} \mu_i = 2 \omega (N_f-N_c)
\end{equation}
In the following we will show how to obtain these identities
by using a standard trick \cite{Spiridonov:2011hf,Benini:2011mf}. 
This correspond to  
derive them by deforming the ones for the 
$USp(2 N_c)$ theories.

Let us end this subsection by briefly commenting on the divergence of the partition function discussed on page 43 of \cite{Aharony:2013kma} for the case of $W=Y$. For the case of $W=Y^2$, we have a different balancing condition from the linear monopole superpotential case.  As a result, we do not have a divergence term that arises from $\Gamma_h(0)$.  In fact, we have checked the aforementioned identities numerically for various parameters $\mu_i$ and found that the results are finite on both sides of the equality.

\subsubsection{Even orthogonal case}

Consider $USp(2N_c)$ with $2N_f+4$ fundamentals and assign 
the masses as follows
\begin{equation}
\mu_{2N_f+1} = 0,\,
\mu_{2N_f+2} = \frac{\omega_1}{2},\,
\mu_{2N_f+3} =  \frac{\omega_2}{2},\,
\mu_{2N_f+4} = \omega\,
\end{equation}
By using the duplication formula \cite{Spiridonov:2011hf}
\begin{equation}
\Gamma_h(2x) = 
\Gamma_h(x)
\Gamma_h\Big(x+\frac{\omega_1}{2}\Big)
\Gamma_h\Big(x+\frac{\omega_2}{2}\Big)
\Gamma_h(x+\omega)
\end{equation}
we have
\begin{equation}
\prod_{a=1}^{N_c} \frac
{\prod_{i=2N_f+1}^{2N_f+4} \Gamma_h(\mu_i\pm x_a)}
{\Gamma_h(\pm 2 x_a)}
=
1
\end{equation}
The partition function of the electric theory becomes
\begin{equation}
\frac{1}{2^{N_c} N_c!}
\int \prod_{a=1}^{N_c} d x_a
\prod_{i=1}^{2 N_f} \Gamma_h(\pm x_a + \mu_i)
\prod_{1\leq a<b \leq N_c} 
\frac{1}{\Gamma_h(\pm x_a \pm x_b)}
\end{equation}
and the balancing condition is 
\begin{equation}
\sum_{i=1}^{2 N_f} \mu_i =  \omega (2 N_f-2N_c+1)
\end{equation}
We can do the same in the dual theory obtaining
\begin{eqnarray}
&&\prod_{1 \leq i<j \leq 2 N_f} \Gamma_h(\mu_i+ \mu_j)
\frac{1}{2^{N_f-N_c-1} (N_f-N_c-1)!}
\int \prod_{a=1}^{N_f- N_c +1} d x_a
\nonumber \\
&&
\prod_{i=1}^{2N_f} \Gamma_h(\pm x_a +\omega- \mu_i)
\prod_{1\leq a<b \leq \widetilde N_c} 
\frac{1}{\Gamma_h(\pm x_a \pm x_b)}
\end{eqnarray}
with two extra pieces
\begin{eqnarray}
\prod_{i=1}^{N_f} 
\Gamma_h(\mu_i) 
\Gamma_h \Big(\mu_i + \frac{\omega_1}{2} \Big) 
\Gamma_h \Big(\mu_i + \frac{\omega_2}{2} \Big) 
\Gamma_h(\mu_i+\omega) 
=
\prod_{i=1}^{N_f}
\Gamma_h (2 \mu_i)
\end{eqnarray}
and
\begin{eqnarray}
\Gamma_h\Big(\frac{\omega_1}{2} \Big)
\Gamma_h\Big(\frac{\omega_2}{2} \Big)
\Gamma_h\Big(\omega+\frac{\omega_1}{2} \Big)
\Gamma_h\Big(\omega+\frac{\omega_2}{2} \Big)
\Gamma_h(\omega)^2=1
\end{eqnarray}
such that the contribution of the meson becomes
\begin{eqnarray}
\prod_{1 \leq i \leq j \leq 2 N_f} \Gamma_h(\mu_i + \mu_j)
\end{eqnarray}
representing the fact that the meson is symmetric in this case.
We have obtained an identity between an $O(2 N_c)$ theory
with $2 N_f$ fundamentals and 
an  $O(2 N_f-2N_c+2)$ theory
with $2 N_f$ fundamentals and a symmetric meson.
The electric and the magnetic monopole superpotentials, compatible
with the balancing condition are quadratic.

\subsubsection{Odd orthogonal case}

In this case we consider $2N_f+2$ fundamentals and fix
\begin{eqnarray}
\mu_{2N_f+1} = \frac{\omega_1}{2},\quad
\mu_{2N_f+2} =  \frac{\omega_2}{2},
\end{eqnarray}

The electric partition function becomes
\begin{eqnarray}
\frac{1}{2^{N_c} N_c!}
\int \prod_{a=1}^{N_c} d x_a
\frac{\prod_{i=1}^{2 N_f} \Gamma_h(\pm x_a+ \mu_i)}{\Gamma_h(\pm x_a)}
\prod_{1\leq a<b \leq N_c} 
\frac{1}{\Gamma_h(\pm x_a \pm x_b)}
\end{eqnarray}
with the balancing condition 
\begin{equation}
\sum_{i=1}^{2 N_f} \mu_i =  \omega (2 N_f-(2N_c+1)+1)
\end{equation}
The dual partition function becomes
\begin{eqnarray}
\!\!\!
\frac{\prod_{1 \leq i<j \leq 2 N_f} \Gamma_h(\mu_i + \mu_j)}{2^{N_f-N_c} (N_f-N_c)!}
\!\!\!
\int
\!\! \prod_{a=1}^{\widetilde N_c} d x_a
\frac{\prod_{i=1}^{2 N_f} \Gamma_h(\pm x_a +\omega- \mu_i)}{\Gamma_h(\pm x_a)}
\!\!\!\!\!\!\!
\prod_{1\leq a<b \leq \widetilde N_c} 
\!\!\!
\frac{1}{\Gamma_h(\pm x_a \pm x_b)}
\end{eqnarray}
with the extra piece
\begin{eqnarray}
\prod_{i=1}^{2N_f}
\Gamma_h \Big(\mu_i + \frac{\omega_1}{2} \Big) 
\Gamma_h \Big(\mu_i + \frac{\omega_2}{2} \Big) 
=
\prod_{i=1}^{2N_f}
\frac
{\Gamma_h (2 \mu_i )}
{\Gamma_h (\mu_i +\omega ) 
\Gamma_h (\mu_i  ) }
=
\prod_{i=1}^{2N_f}
\frac
{\Gamma_h (2 \mu_i )
\Gamma_h (\widetilde \mu_i  ) }
{
\Gamma_h (\mu_i ) }
\end{eqnarray}
By distributing properly the terms on the RHS and LHS we 
arrive at the expected relation between the orthogonal 
theories with odd rank and quadratic superpotential

\subsubsection{Flowing to Aharony in the orthogonal case}

We can engineer the flow from the $W= Y^2$ duality to Aharony duality in the orthogonal case as well. This is a further check of the proposed duality.
In the electric case we consider $2(N_f+1)$ fundamentals and assign the large real masses as $\mu_{2N_f+1}= \frac{\eta}{2}+s$ and
$\mu_{2N_f+2} = \frac{\eta}{2}-s$.
In the magnetic theory we must consider an higgsing of the dual gauge group as well, giving raise to an $O(N_f-N_c+2) \times U(1)$ theory.
The $O(N_f-N_c+2)$ theory has $N_f$ fundamentals and superpotential
$W = M q q$, where the meson $M$ is a symmetric tensor with dimension $N_f (2N_f+1)$. There is an extra massless singlet $H$ in this theory, coming from the original meson with dimension $(N_f+1)(2N_f+3)$.
This interact with the two chiral multiplets, having opposite charge under the $U(1)$ sector. We refer to them as $p$ and $\tilde p$, such that the superpotential of this sector is $W_{U(1)} = H p \tilde p$.
This $U(1)$ sector has an FI term constrained by the choice of real masses. If we dualize this sector this is dual to three, the meson 
$X =  p \tilde p$, and the monopoles $Y$ and $Z$.
The fields $H$ and $X$ are massive because of the superpotential.
The duality map fixes $Y$ to have vanishing real mass and R-charge equal to 1. This corresponds to have a massive field.
The other field $Z$ has the same quantum numbers of the electric
monopole and it is compatible with the dual superpotential of the Aharony
dual model.
 This structure can be reproduced on the partition function by distinguishing the even and odd cases as before.


\section{Matching the Hilbert series} \label{sec:HSquad}
In this section, we compute the Hilbert series for theories with quadratic monopole superpotentials.  

\subsection{Models with unitary gauge groups}
\subsubsection{$W = (X^+)^2 +  (X^-)^2 $} \label{sec:HSXpsqXmsq}
In this section, we consider the duality presented in section \ref{sec:XpsqXmsq}.  

Let us first analyse theory $A$. The residual theory is the whole $U(N_c)$ with $N_f$ flavours, with the $R$-charge $Q$ and $\tilde{Q}$ is given by \eref{RXpsqXmsq}.  The Hilbert series of theory $A$ is thus the mesonic Hilbert series of $U(N_c)$ with $N_f$ flavours
\be
\begin{split}
H^{(A)} &= H^U_{N_c,N_f}(t,\vec{u},\vec{v};R)~. \\
\end{split}
\ee
where the expression for $H^U_{N_c,N_f}$ is given in \eref{mesHU} and the $R$-charge $R$ of the quarks and antiquarks are given by \eref{RXpsqXmsq}.  There is no fugacity $y$ because the axial symmetry is broken by the monopole superpotential.

Let us now analyse theory $B${, which is a $U(N_f-N_c)$ gauge theory with $N_f$ flavours $q$ and $q$, singlets $M$ and the superpotential $W=M q \tilde{q} + (\hat{X}^+)^2 + (\hat{X}^-)^2$, where $\hat{X}^\pm$ are the basic monopole operators in this theory.  In order to study the moduli space of this theory, we find that it is convenient to use the Aharony duality of theory $B$ as a tool to study.  Such a duality gives the following theory $B'$:}
\paragraph{Theory $B'$:} $U(N_c)$ gauge theory with $N_f$ flavours $\mathfrak{q}$ and $\tilde{\mathfrak{q}}$, two singlets $\hat{X}^\pm$ and superpotential
\be
W_{B'}=  \hat{X}^+V^- +\hat{X}^-V^++ (\hat{X}^+)^2+ (\hat{X}^-)^2~,
\ee
where $V^\pm$ are the basic monopoles in theory $B'$. { We emphasise that the basic monopole operators of theory $B$ are identified the singlets $\hat{X}^\pm$ of theory $B'$ under the duality map.  From the superpotential $W_{B'}$, we see that the singlets $\hat{X}^\pm$ are massive and can be integrated out.  Upon substituting the $F$-terms $\partial_{\hat{X}^\pm} W_{B'}=0$, which give $V^{\mp}=-2 \hat{X}^\pm$, back into $W_{B'}$, we obtain $-\frac{1}{4}[(V^+)^2+(V^-)^2]$, which is the superpotential of theory $A$ (up to a factor of $-1/4$).  This provides a consistency check of the proposed duality.  As an immediate consequence, the Hilbert series of theory $B$, which is dual to theory $B'$ and is thus identical to that of theory $A$. 

Moreover, since the singlets $\hat{X}^\pm$ are massive theory $B'$ due to the term $(\hat{X}^\pm)^2$ in $W_{B'}$, by the duality map, we expect the basic monopole operators of theory $B$ to be massive and the Coulomb branch of theory $B$ to be lifted.  This is consistent with our proposal that the presence of the quadratic monopole superpotential terms lead to the lift of the Coulomb branch.}

\subsubsection{$W = (X^-)^2$} \label{sec:XmsqHS}
Let us now discuss about the duality presented in section \ref{sec:Xmsq}.

\subsubsection*{Theory $A$}
Let us consider theory $A$.  We mix the $U(1)_{T'}$ symmetry with the $R$-symmetry such that the $R-$charges of the $X^\pm$ are given by
\bea
&R[X^-]=N_f(1-R)-(N_c-1) \\
&R[X^+]+\alpha=N_f(1-R)-(N_c-1),
\eea
where
\be
R = \frac{1}{2}(R[Q]+R[\tilde Q])~.
\ee
and $\alpha$ parametrizes the mixing of the UV $R$ charge and the $U(1)_{T'}$ charge of the monopole operators.
Imposing the marginality of the superpotential we get $R[X^-]=1$, we have
\be
N_f(1-R)-(N_c-1)=1~,
\ee
from which we get $R$
\be \label{RtheoryA}
R=\frac{N_f -N_c}{N_f}.
\ee
Also,
\be
R[X^+] = 1- \alpha~.
\ee

Now let us turn our attention to the Hilbert series. Considering the $F$-term $\partial_{X^-} W_A=0$, we see that $X^-=0$ in the chiral ring.  Denoting the magnetic flux by $(m_1,0,\ldots, 0,m_{N_c})$, with $m_1\geq0\geq m_{N_c} =0$, we have 2 possible cases:
\bi
\item {\bf $m_1=0=m_N$:} with residual theory $U(N_c)$ with $N_f$ flavours. The Hilbert series 
is 
\be 
	H_{I}^{(A)}(t, \vec u, \vec v, z; R)=H^U_{N_c, N_f}(t, \vec u, \vec v,z; R).
\ee
where $z$ is the fugacity for the $U(1)_{T'}$ symmetry.
\item {\bf $m_1>0=m_N$:} with residual theory $U(N_c-1)$ with $N_f$ flavours. The Hilbert series is given by
\be
\begin{split}
	H_{II}^{(A)}(t, \vec u, \vec v,z ; R) &=\sum_{m_1=1}^{+\infty} t^{R[X^+](m_1)} z^{T'[X^+](m_1)} H^U_{N_c-1, N_f}(t, \vec u, \vec v, z; R) \\
	&= \frac{t^P z^x}{1-t^P z^x} H^U_{N_c-1, N_f}(t, \vec u, \vec v, z; R) ~,
\end{split}
\ee
where we assign the fugacity $z^x$ for the $U(1)_T'$ symmetry of $X^+$, and
\be
P = 1-\alpha~.
\ee
\ei
Thus, the Hilbert series of theory $A$ reads
\be
\begin{split}
	H^{(A)}(t, \vec u, \vec v, z; R)&=H^U_{N_c, N_f}(t, \vec u, \vec v, z; R) + \left(\frac{t^{P} z^x}{1-t^{P} z^x} \right) H^U_{N_c-1, N_f}(t, \vec u, \vec v, z; R)~. \\
\end{split}
\ee
\subsubsection*{Theory $B$}
Let us consider now the theory $B$, which is an $U(N_f-N_c)$ gauge theory with $N_f$ flavours $q$ and $\tilde q$, $N_f^2$ singlets 
$\hat M$, singlet $S^+$ and superpotential $W=\hat{M} \tilde{q} q + (\hat X^+)^2 + S^+ \hat X^- $, where $\hat X^{\pm}$ are the basic monopole operators in theory $B$.  To analyse the moduli space, it is convenient focus on the Aharony dual of theory $B$, which is given by the following theory $B'$.

%
\paragraph{Theory $B'$:} $U(N_c)$ with $N_f$ flavours and singlets  $\hat X^\pm$ and $S^+$ with superpotential
\be W_{B'}=\hat X^+ V^- + \hat X^- V^+ + (\hat X^+)^2 + \hat X^- S^+ ~.\ee 
where $V^\pm$ are the basic monopole operators of theory $B'$. { The singlets $\hat X^\pm$ in theory $B'$ are mapped to the basic monopole operators in theory $B$ under the duality.}

We see that the singlet $\hat{X}^+$ is massive and can be integrated out. The $F$-terms with respect to the singlets are
\bea
	&\d_{\hat X^-} W_{B'}=0 \quad \implies \quad V^+=-S^+, \label{VpSp} \\
	&\d_{S^+} W_{B'}=0 \quad \implies \quad \hat X^-=0, \\
	&\d_{\hat X^+} W_{B'}=0 \quad \implies \quad V^-=-2 \hat{X}^+.
\eea 
{ Plugging these equations into $W_{B'}$, we eliminate $S^+$, $\hat{X}^\pm$ and obtain
\be
	W_{B'}=-\frac{1}{4} (V^-)^2~,
\ee
which is the superpotential of theory $A$ (up to a factor of $-1/4$).  This provides a consistency check of the proposed duality.  As an immediate consequence, the Hilbert series of theory $B$, which is dual to theory $B'$ and is thus identical to that of theory $A$. 
Since the singlet $\hat{X}^+$ are massive theory $B'$, by the duality map, we expect the basic monopole operators of theory $B$ to be massive and the Coulomb branch of theory $B$ to be lifted.  This is consistent with our proposal that the presence of the quadratic monopole superpotential terms lead to the lift of the Coulomb branch.}

\subsubsection{$W = (X^-)^2$, chiral flavours and CS terms} \label{sec:XmsqCS}
We now consider the duality involving chiral flavours and CS terms presented in section \ref{sec:XmsqchflvCS}.

The analysis of theory $A$ is very similar to section \ref{sec:XmCS}.  The monopole operator $X^-$ vanishes in the chiral ring { due to the quadratic term $(X^-)^2$ in the superpotential}, and the Coulomb branch generated by $X^+$ is lifted due to the non-zero CS level.  The Hilbert series of theory $A$ is given by \eref{HSABBP} with the $R$-charge $R$ given by \eref{RchargeXmsqCS}.

By the similar argument to section \ref{sec:XmCS},  the moduli space of theory $B$ is generated by $M$ subject to the quantum condition that $\mathrm{rank}(M) \leq N_c$.  In order to prove this, we turn on a VEV of $M$ with rank $N_c+p$ with $p>0$ and obtain the how energy theory of $U(N_f-N_c)_{-k/2}$ gauge theory with $(N_f'=N_f-N_c-p, \,\, N_a'=N_f-k-N_c-p)$ fund/antifund.   We then use the BCC duality to obtain the dual theory.  Since the dual gauge group is the unitary group of rank $N_f'-(N_f-N_c) = -p$, we see that for $p>0$, supersymmetry is broken and there is no supersymmetric vacuum.

\subsection{Models with symplectic gauge groups} \label{sec:HSsympY2}
We now consider the duality involving symplectic gauge groups presented in section \ref{sec:sympY2}.

In theory $A$ the Coulomb branch is completely lifted { due to the quadratic term $Y^2$ in the superpotential}. The Hilbert series of the theory is
thus the mesonic Hilbert series
\be
	H^{(A)}=H^{USp}_{2N_c, 2N_f}(t, \vec x; R),
\ee
where the expression for $H^{USp}_{2N_c, 2N_f}$ is given by \eref{HUSp} and the $R-$charge $R$ of the fundamentals is given by \eref{RmonosympY2}.  There is no fugacity $y$ because the $U(1)$ axial symmetry is broken.

Let us consider now theory $B$. It is convenient to study its Aharony duality.  Let us call the latter $B'$. This is a $USp(2N_c)$ gauge theory with a singlet $\hat{Y}$ and { superpotential 
$W'=  \hat{Y} \hat Y' +\hat{Y}^2$
where $\hat Y'$ is now the basic monopole of this theory. Here $\hat{Y}$ is massive and can be integrated out. The $F$-terms of the singlet give $\hat Y'=-2\hat{Y}$.  Substituting back in $W'$, we obtain a superpotential $W'=-\frac{1}{4}\hat{Y}'^2$.  This is the superpotential of theory $A$ (up to a factor of $-1/4$).  This provides a consistency check of the proposed duality.  As an immediate consequence, the Hilbert series of theory $B$, which is dual to theory $B'$ and is thus identical to that of theory $A$.   Since the singlet $\hat{Y}$ are massive theory $B'$, by the duality map, we expect the basic monopole operators of theory $B$ to be massive and the Coulomb branch of theory $B$ to be lifted.  This is consistent with our proposal that the presence of the quadratic monopole superpotential terms lead to the lift of the Coulomb branch.}

\subsection{Models with orthogonal gauge groups} \label{sec:HSorthY2}
We now consider the duality involving orthogonal gauge groups presented in section \ref{sec:orthY2}.

Let us first analyse theory $A$.  The Coulomb branch is completely lifted due to the $F$-term $\partial_Y W=0$, so the Hilbert series of the theory is
the mesonic Hilbert series
\be
	H^{(A)}(t, \vec x; R)=H^{O}_{N_c, N_f}(t,  \vec x; R),
\ee
where the expression for $H^{O}_{N_c, N_f}$ is given by \eref{mesO} and $R$ is the $R$-charge of the quarks given by \eref{RmonoorthY2}.

Let us now turn to theory $B$.  As before, it is convenient to consider the Aharony dual of this theory $B$.  { Let us call this theory $B'$. It is an $O(N_c)$ gauge theory with a singlet $\hat Y$ and superpotential $W'=  \hat{Y}' \hat Y+\hat Y^2$,
where $\hat{Y}'$ is the basic monopole of this theory. Here $\hat Y$ is massive and can be integrated out. The $F$-term of the singlet gives  $\hat{Y}'=-2\hat Y$. Substituting 
back in $W'$ we obtain a superpotential $W'=- \frac{1}{4}\hat Y'^2$, which is the superpotential of theory $A$ (up to a factor of $-1/4$). As in the previous subsection, this provides the consistency of the duality.  The Hilbert series of theory $B$, which is dual to theory $B'$ and is thus identical to that of theory $A$.}

For the duality involving the special orthogonal gauge groups, the analysis is very similar.  For theory $A'$, the Hilbert series is 
\be
	H^{(A')}(t,  \vec x; R)=H^{SO}_{N_c, N_f}(t,  \vec x; R),
\ee
where $H^{SO}_{N_c, N_f}$ is given by \eref{HSO}.  The moduli space of theory $B'$ can be analysed by using the ARSW duality in the same way as the above.

\section{Further developments}
\label{sec-conc}
Let us conclude this paper by discussing an obstruction to the duality for theories with quartic powers of the monopole superpotential. We then proceed to discuss open problems and other research directions that we leave for future work.

\subsection{The quartic monopole superpotential}
Let us consider the following pairs of theories:
\paragraph{Theory $A$:} $U(N_c)$ gauge theory with $N_f$ flavours $Q$ and $\tilde{Q}$, and superpotential
\be
W_A= (X^+)^4+(X^-)^4~.
\ee
\paragraph{Theory $B$:} $U(N_f-N_c+1)$ gauge theory with $N_f$ flavours $q$ and $\tilde{q}$, $N_f^2$ singlets $M$ and superpotential 
\be W_B= M \tilde{q} q+ (\hat{X}^+)^4 + (\hat{X}^-)^4~.
\ee
\\
The $R$-charge $R_A$ of the quarks and antiquarks in theory $A$ is fixed by the monopole superpotential:
\be
R[X^\pm] = \frac{1}{2} = N_f(1-R_A) - (N_c-1) \quad \Rightarrow \quad R_A= \frac{2N_f-2N_c+1}{2N_f}~.
\ee
A similar computation show that the $R$-charge $R_B$ of the quarks and antiquarks in theory $B$ is
\be
R[\hat{X}^\pm] = \frac{1}{2} = N_f(1-R_B) - (N_f-N_c+1-1) \quad \Rightarrow \quad R_B= \frac{2N_c-1}{2N_f}~.
\ee
Indeed, the $R$-charge of $M$ in theory $B$ is $R[M]=2-2R_B = \frac{2N_f-2N_c+1}{N_f}$.  This is equal to the $R$-charge of the mesons in theory $A$, where latter is $2R_A = \frac{2N_f-2N_c+1}{N_f}$.  Thus, theories $A$ and $B$ have a chance to be dual to each other.  However a further analysis shows that this is not the case.

In theory $A$, { we use the similar argument as in section \ref{sec:effectquad} that the presence of $(X^\pm)^4$ terms in the superpotential lifts the Coulomb branch.} The residual theory is thus a $U(N_c)$ gauge theory with $N_f$ flavours.  The meson matrix thus has a maximum rank of $N_c$.  

On the other hand, in theory $B$, if we give a VEV of rank $N_c$ to the singlet $M$, which is dual to the meson in theory $A$, the low energy effective field theory is $U(N_f-N_c+1)$ gauge theory with $N_f-N_c$ flavours. According to \cite{Aharony:1997bx}, this theory is described by the monopole operators $\hat{X}^\pm$ and the mesons $q \tilde{q}$ satifying $\hat{X}^+ \hat{X}^- \det(q \tilde{q}) =1$.\footnote{The $R$-charge of $\hat{X}^\pm$ in this effective theory is $R[\hat{X}^\pm]= (N_f-N_c)(1-R) -(N_f-N_c+1-1)$, where $R=R[q]=R[\tilde{q}]$ and $R[\det(q \tilde{q})] = 2(N_f-N_c) R$.  Thus, $\hat{X}^+ \hat{X}^- \det(q \tilde{q})$ has $R$-charge $0$.} The effective superpotential of theory $B$ is therefore
\be
W_B' = (\hat{X}^+)^4 + (\hat{X}^-)^4 + \lambda ( \hat{X}^+ \hat{X}^- \det(q \tilde{q})-1 )~,
\ee
where $\lambda$ is a Lagrange multiplier.
Using the equations of motion, we find that the effective superpotential becomes $W_B' \sim (\det(q \tilde{q}))^{-2}$.  Thus, we have runaway vacua.  This does not match with theory $A$, where we have the mesonic branch of $U(N_c)$ gauge theory with $N_f$ flavours as the moduli space of vacua.  Thus, the duality fails.

This argument can be used to show that the following pairs of theories {\bf cannot} be dual to each other:
\paragraph{Theory $A$:} $O(N_c)$ gauge theory with $N_f$ flavours, and superpotential $W_A = Y^4$.
\paragraph{Theory $B$:} $O(N_f-N_c+3)$ gauge theory with $N_f$ flavours $q$, $N_f(2N_f+1)$ singlets $M$ and superpotential $W_B= M q q+ \hat{Y}^4$.
\\~\\
The $R$-charge of the mesons in theory $A$ is equal to that of $M$ in theory $B$; they are equal to $\frac{2N_f-2N_c+3}{N_f}$.
The maximum rank of the meson matrix in theory $A$ is $N_c$.  Applying the BCC duality to theory $B$, we obtain the dual gauge group being $O(N_c-1)$. The maximum rank of the meson matrix in the latter is $N_c-1$, which is not compatible with that in theory $A$.

Let us finally consider the following pairs of theories:
\paragraph{Theory $A$:} $U(N_c)$ gauge theory with $N_f$ flavours, and superpotential $W_A = (X^-)^4$.
\paragraph{Theory $B$:} $U(N_f-N_c+x)$ gauge theory with $N_f$ flavours $q, \, \tilde{q}$, $N_f^2$ singlets $M$, a singlet $S^+$ and superpotential $W_B= M q \tilde{q}+ S^+ \hat{X}^- + (\hat{X}^+)^4$.
\\~\\
As before, the maximum rank of the meson matrix of theory $A$ is $N_c$.  Let us consider the Aharony dual of theory $B$ { and call this theory $B'$. The latter is} a $U(N_c-x)$ gauge theory with $N_f$ flavours, singlets $S^+,\, \hat{X}^+, \, \hat{X}^-$, $N_f^2$ singlets $M$ and superpotential $W_{B'}= \hat{X}^+ V^-+\hat{X}^- V^+ +S^+ \hat{X}^- + (\hat{X}^+)^4$, { where $V^\pm$ are the basic monopole operators of theory $B'$}.  Using the $F$-terms $\partial_{\hat{X}^\pm} W_{B'}=0$ and $\partial_{S^+} W_{B'}=0$ and substituting back to $W_{B'}$, we obtain the effective superpotential $W_{B'} \sim (V^-)^{4/3}$. {  Observe that, for this model, we do not recover the superpotential of theory $A$ under this procedure, as the theories discussed earlier in section \ref{sec:HSquad}.  Let us, nevertheless, proceed further.  This effective superpotential sets the $R$-charge of $V^-$ to be $R[V^-] = \frac{3}{2}$.  The $R$-charges of $X^-$ of theory $A$ and $V^-$ of theory $B'$ are given by
\bea
\frac{1}{2}-\beta &= R[X^-] = N_f(1-R_A)-(N_c-1)~,\\
\frac{3}{2}-\beta &= R[V^-] = N_f(1-R_{B'})-(N_f-N_c+x-1)~,
\eea
where $\beta$ parametrises the mixing between $U(1)_R$ and $U(1)_{T'}$; and $R_A$ and $R_{B'}$ are the $R$-charges of quarks and antiquarks in theories $A$ and $B'$, respectively. Since the mesons in theory $A$ are mapped to the mesons in theory $B'$, we also require that
\be
R_A = R_{B'}~.
\ee
Solving these three equations, we find that
\be
x= 2N_c - N_f -1~.
\ee
Thus theory $B'$ has gauge group $U(N_c-1)$, and the maximum rank of the meson matrix is $N_c-1$.  This is in contradiction with theory $A$.
}

\subsection{Future directions}
Let us discuss some other interesting lines of research that we leave
for future analysis.

An important aspects that we did not discuss in the paper is related to 
estimation of the conformal window.
This is indeed possibile that 
some singlets hit the bound of unitarity when maximizing the free energy.
This bounds corresponds to the failure of the inequality  
$\Delta_{Singlet} > \Delta_{Free} = \frac{1}{2}$.
In such cases we are in presence of accidental symmetries that need to be cured by applying the procedure of \cite{Kutasov:2003iy} 
(see also \cite{Agarwal:2012wd} for a 3d version of this procedure).
This can modify the dualities and it should be interesting to have a complete
understanding of the conformal window along the lines of \cite{Safdi:2012re}.
A related analysis consists of finding UV complete models that flow to the ones with quadratic power monopole superpotentials in the IR, of the type discussed here. Similar discussions appeared in \cite{Benini:2017dud} and it would be interesting to adapt such analysis to our cases.

Another aspects that we did not discuss is related to the global aspects of the dualities with orthogonal gauge groups in presence of quadratic power monopole superpotentials.
In such cases one should follow the discussion of \cite{Aharony:2013kma}
and distinguish $O(N)_\pm$ and Spin$(N)$ cases.
Here we restricted to the duality between $O(N)_+$ groups.

Furthermore we did not discuss possible dualities between $SU(N)$ gauge groups and quadratic power monopole superpotentals. The existence of a 
duality for the $USp(2)=SU(2)$ case looks a good starting point for the existence of such duality, but we have not been able to provide a general
behaviour for such a case. 

Another interesting problem consists of the brane interpretation of 
the quadratic monopole superpotential. Naively one can think to this 
superpotential as arising as in the linear case \cite{Amariti:2015yea}, i.e. by placing a 4D theory
on a circle, T-dualizing, possibly moving some D-brane (D3 and or D5) 
along the circle and then adding a D1 brane between two stacks of D3 
branes separated along the compact direction.
These D1 branes represent the monopole superpotential, KK monopole in the case of real gauge groups
\cite{Amariti:2015mva}  and linear monopoles for unitary groups 
\cite{Amariti:2017gsm}.
A similar construction may be engineered for the quadratic monopoles, where the higher power can be for example engineered by multiple stacks of D1 branes.
It should be interesting to check this or similar constructions.

\acknowledgments
We thank Francesco Benini, Sergio Benvenuti, Luca Cassia, Stefano Cremonesi, Simone Giacomelli, Gabriele Lo Monaco, and Sara Pasquetti for valuable discussions.  A.A. and N.M. gratefully acknowledge the Simons Center for Geometry and Physics for the Simons Summer Workshop 2016,  where this work was initiated.  A.A. and N.M. are also indebted to the Galileo Galilei Institute for Theoretical Physics for the hospitality during the Workshop on Supersymmetric Quantum Field Theories in the Non-perturbative Regime, where substantial progress on this work has been made.  We thank INFN for support during the completion of this work.

\bibliographystyle{ytphys}
\bibliography{ref}

\end{document}